\documentclass[prb,twocolumn,aps,showpacs,amsmath,amssymb]{revtex4-1}

\usepackage{bm}
\usepackage[colorlinks=true, citecolor=blue, urlcolor=blue]{hyperref}\usepackage{epsf,graphics,graphicx}
\usepackage{amsmath}
\usepackage{latexsym}
\usepackage{amssymb}
\usepackage{graphicx}
\usepackage{color}
\usepackage{soul}
\usepackage[dvipsnames]{xcolor}
\setstcolor{red}
\usepackage{ulem}
\usepackage{cancel}
\usepackage{tikz}
\usepackage[export]{adjustbox}
\usepackage{subcaption}
\usepackage{braket}

\def\be{\begin{equation}}
\def\ee{\end{equation}}
\def\ba{\begin{eqnarray}}
\def\ea{\end{eqnarray}}

\newcommand{\bi}{\bibitem}

\begin{document}
\title{\textcolor{black}{
Thermo-electric transport properties of Floquet multi-Weyl Semimetals}}
\author{Tanay Nag$^1$, Anirudha Menon$^2$, and Banasri Basu$^3$} 
\affiliation{$^1$ SISSA, via Bonomea 265, 34136 Trieste, Italy}
\affiliation{$^2$ Department of Physics, University of California, Davis, California 95616, USA}
\affiliation{$^3$ Physics and Applied Mathematics Unit, Indian Statistical Institute, Kolkata 700108, India}

\begin{abstract}
We discuss the  circularly polarized light (of amplitude $A_0$ and frequency $\omega$)
driven thermo-electric transport properties of type-I and type-II multi-Weyl semimetals (mWSMs) in the high frequency limit. Considering the low energy model, we employ the Floquet-Kubo formalism to compute the thermal Hall and Nernst conductivities for both  types of mWSMs. We show that the anisotropic nature of the dispersion for arbitrary integer monopole charge $n>1$ plays an important role in  determining the effective Fermi surface behavior; interestingly, one can observe momentum dependent corrections in Floquet mWSMs in addition to momentum independent contribution as observed for Floquet single  WSMs. Apart from the non-trivial tuning of the Weyl node position $\pm Q \to \pm Q- A_0^{2n}/\omega$, our study reveals that the momentum independent  terms result in leading order contribution in the conductivity tensor. This has the form of $n$ times the single WSMs results with effective chemical potential $\mu \to \mu -A_0^{2n}/\omega$. On the other hand, momentum dependent corrections lead to sub-leading order terms which are algebraic function of $\mu$ and are present for $n>1$. Remarkably, this analysis further allows us to distinguish type-I mWSMs from their type-II counterparts. For type-II mWSMs, we find that the transport coefficients for $n\geq 2$ exhibit algebraic dependence on the momentum cutoff in addition to the weak logarithmic dependence as noticed for $n=1$ WSMs. We demonstrate the variation and qualitative differences of transport coefficients between type-I and type-II mWSM as a function of external driving parameter $\omega$. 

\end{abstract}

\maketitle

\section{Introduction}
\label{s1}

Recent years have witnessed Weyl Semimetals (WSMs) as a focus of research  attraction due to their 
exotic properties. The upsurge of recent attention on this new class of quantum materials is due to its unusual Fermi arc surface states and chiral anomaly that is intimately related to topological order \cite{hasan17, armitage18}. In WSMs, the bulk band gap closes at an even number of discrete points in the Brillouin zone. These special gap closing points, protected by some crystalline symmetry, are referred as Weyl nodes \cite{wan11};  Weyl nodes act as a monopoles or anti-monopoles of Berry curvature characterized by integer monopole charge $n$. \textcolor{black}
 { Two Weyl nodes of different chirality are located at different momenta when the system breaks the time reversal symmetry; four Weyl nodes are noticed in general for system with broken inversion symmetry only \cite{trivedi17,armitage18}. Moreover, inversion breaking may also lead to the energy separated Weyl points 
while time reversal symmetry breaking can result in  Weyl points at 
same energy \cite{zyuzin12a,dey20}.} The existence of  Fermi arc surface states,  chiral-anomaly related negative magnetoresistance, and the quantum anomalous Hall effect are the direct consequences of the topological nature of WSMs\cite{zyuzin12,son13,burkov15}. As compared to the conventional WSMs with $n = 1$, reported in TaP, TaAs, NbAs \cite{xu15a,lv15a,lv15b}, it has been recently shown that $n$ can be generically greater than one, with the crystalline symmetries bounding its maximum value to three \cite{xu11,fang12,yang14}. 
These are called multi WSMs (mWSMs); interestingly, the single-WSM with $n=1$ can be considered as 3D analogue of graphene whereas the double WSM (triple WSM) with $n=2$ ($n=3$) can be
represented as 3D counterparts of bilayer (ABC-stacked trilayer) graphene \cite{falko06,peres06,macdonald08}. Close to the Weyl points, mWSMs host low-energy quasiparticles with the dispersion which is, in general, linear only in one direction leading to  anomalous features in the transport  properties \cite{park17, gorbar17, dantas18,lepori18,nag18, sinha19, dantas19}.\\

An ideal WSM has a conical spectrum and a point-like Fermi surface at the Weyl point.
An interesting situation arises when large tilting of the Weyl cones results in a Lifshitz   transition. This leads to a new class of materials called type-II  WSMs, where the  Fermi  surface  is no  longer point-like \cite{volovik14,xu15,soluyanov15, yan17,armitage18,menon18}. The existence of type-II WSM has been experimentally demonstrated \cite{li17,kimura19} while theoretical prediction shows that a type-II WSM can be engineered  by applying strain or chemical doping to the original type-I WSM \cite{trescher15}. The type-II WSM phase is characterized by a different class of Weyl fermions manifesting the violation of Lorentz symmetry. Type-II WSMs can yield intriguing electronic transport properties due to a markedly different density of  states  at the Fermi level \cite{yu16,udagawa16,fei17,lv17}. In addition to electric transport, thermal responses also carry signatures  of the exotic physics of WSMs which have been studied theoretically \cite{landsteiner14, sharma16,lucas16,chen16a} and experimentally  \cite{hirschberger16,chen16,watzman18,stocker17}. At the same time, optical conductivity of WSMs have been extensively studied along with other characteristic signatures \cite{tabert16,tabert16a,mukherjee17,mukherjee18,sonowal19,das19}. While a lot of progress has been made experimentally and theoretically in investigating $n=1$ type-I and type-II WSMs, the experimental discovery of mWSMs with $n \geq 2$ is yet to be made; however, using density functional theory calculations some materials are conjectured to host Weyl nodes with monopole charges $n=2,3$ \cite{xu11,bernevig12,hasan16,zunger17}. This further  motivates the theoretical search for finding additional tools to identify these exotic phases with higher monopole charge. For example, mWSMs, in general, can exhibit a smoothly deformed conical spectrum  and a point-like Fermi surface at the Weyl point. Interestingly, for type-II mWSMs, these features are expected to change and might lead to distinct transport characteristics as compared to type-I mWSMs. \\

On the other hand, periodically driven Floquet systems, where the static Hamiltonian is
perturbed with a time-periodic drive, have attracted a lot of interest recently. Floquet 
systems can host unique phases which have no counterparts in equilibrium systems, such as anomalous Floquet topological phases \cite{oka09, kitagawa10, rudner13, nathan15, titum16, po16, mukherjee17,kar18}, dynamical freezing \cite{das10}, many-body energy localization \cite{alessio13}, dynamical localisation\cite{nag14,agarwala16}, Floquet higher order topological phases \cite{nag19}, and dynamical generation of edge Majorana modes\cite{thakurathi13}. It has been shown that circularly polarized light can be employed to switch between Weyl semimetal, Dirac semimetal and topological insulator phases in a prototypical three-dimensional (3D) Dirac material, Na$_3$Bi \cite{hubener17}. Furthermore, the D.C. transport is expected to be drastically modified under such irradiation \cite{kitagawa10,gu11}. Interestingly, linearly polarized light can lead a band insulator to a WSM phase where the relative separation of Weyl points can be controlled \cite{zhang16}; similarly, circularly polarized light drives a nodal line semimetals into Weyl semimetals \cite{yan16}. In the high frequency driving limit, the system does not absorb energy via electronic transitions, resulting in a non-equilibrium steady state. In this limit, an effective static Sambe space Hamiltonian picture successfully describes the non-trivial outcomes \cite{goldman14,mikami16}.\\

Given the background on the generation and optical manipulation of Weyl nodes, our aim here is to study the thermo-electric transport properties of  mWSMs when it is driven by a circularly polarized source in the high frequency limit.  \textcolor{black}{The irradiation can  act differently depending on whether the underlying static system obeys or breaks the time reversal symmetry \cite{oka09,zhang16}. For example, the irradiated graphene, which is intrinsically time reversal invariant, becomes topologically gapped whereas light induced WSM, which can be intrinsically time reversal broken, remains gapless with renormalized Weyl node position. 
} A recent study using non-equilibrium Kubo formalism have revealed that the  thermo-electric response of type-I WSMs can be distinguished from type-II WSMs under the application of light \cite{menon18}. \textcolor{black}{ One can hence note that in mWSMs, the anisotropic dispersion may lead to unusual outcome as compared to the single WSMs and graphene.
It is natural to ask} the question ``how do the thermal Hall conductivity and Nernst conductivity of the type-I phase differ qualitatively and quantitatively from the type-II phase when the underlying Weyl Hamiltonian supports higher topological charge $n>1$?"
\textcolor{black}{This paper is an attempt to answer the above question; in particular, our work yields a general framework for Weyl systems from which the single Weyl results can be obtained directly. } \\

We find for $n>1~(n=1)$ that the position of  the Weyl point can be tuned in a non-trivial (trivial) manner and {\color{black} the Fermi surface gets renormalized with both momentum dependent and independent (only momentum independent) terms.} These additional interesting features in the $n>1$ case heavily influence the subsequent transport properties. The momentum independent term gives $n$ times the single Weyl results for conductivity tensor in its leading order while the dependent terms can lead to sub-leading correction in the conductivity tensor. Our study further suggests that the vacuum contribution becomes cut-off dependent, unlike the $n=1$ case, due to the coupling of the $U(1)$ gauge field to the anisotropic dispersion
that contains higher momentum modes. In addition to the logarithmic cut-off dependence in Fermi surface contribution for type-II $n=1$ WSM, we find a strong algebraic cut-off dependence for $n \geq 2$. Interestingly, the Fermi surface contribution for type-I mWSM continues to show a cut-off independent response, similar to the observation for $n=1$ mWSMs. Type-I mWSMs behave in a dissimilar manner as compared to type-II mWSMs, as a function of the chemical potential; this is very clearly visible when the Nernst conductivity is investigated. We wish to note that the results presented here pertain to the minimal model and further investigation is needed to verify our claims.\\

The paper is organized as follows: Sec.~\ref{eff_ham} discusses the equilibrium and non-equilibrium low-energy model Hamiltonian and compares it with the single WSM case. We then study the Berry curvature and anomalous Hall conductivity in detail in Sec.~\ref{Berry_curvature}. Next in Sec.~\ref{sigma_xy}, we present our analytical results for optical conductivity using Floquet-Kubo formalism and extensively analyze the vacuum and fermi surface contribution. In Sec.~\ref{DR} we pictorially represent the distinctive behaviors of type-I and type-II mWSMs, and discuss the underlying physics. We present our conclusions in Sec.~\ref{conclusion}.


\section{Effective Floquet Hamiltonian}
\label{eff_ham}
The low energy Hamiltonian for a multinode WSM of monopole charge $n$ near each Weyl point is given by\cite{model}
\begin{equation}
 \label{h1}
H_{\mathbf k}^s=\hbar  C_s( k_z-sQ) + s \hbar \alpha_n {\boldsymbol{\sigma}} \cdot ({\mathbf n}_{\mathbf k}- s {\mathbf e}).
\end{equation}
\textcolor{black}{The lattice model for mWSMs can be shown to reduce in the  
 above low energy model \cite{supple} }.
Here, $s=\pm$ indicates the chirality of nodes, 
${\mathbf n}_{\mathbf k}=[k_\perp^n \cos(n \phi_{\mathbf k}), k_\perp^n \sin (n \phi_{\mathbf k}), \frac{v k_z}{\alpha_n }] $. ${\mathbf e}=(0,0,Q)$, and $2 Q$ is the separation between two Weyl nodes. \textcolor{black}{
$\boldsymbol{\sigma} = [\sigma_x,\sigma_y,\sigma_z]$ is the vectorized Pauli matrix}, and $\alpha_n$ is the mWSM coupling which reduces to the Fermi velocity $v$ when $n=1$. We define the $x-y$ plane azimuthal angle $\phi_{\mathbf k}={\rm \arctan}(\frac{k_y}{k_x})$, and the in-plane momentum $k_{\perp}=\sqrt{k_x^2+k_y^2}$. The Hamiltonian (\ref{h1}) represents the two Weyl nodes $(0,0,\pm Q)$, located at the same energy and separated by a distance $2Q$, while $C_s$ indicates the tilt parameter associated with $s$ Weyl node. Type-I mWSMs corresponds to $|C_s|/v\ll 1$ while for type-II mWSMs we have $|C_s|/v \gg 1$. We restrict to the inversion symmetric tilt given by  $s C_s=C$.
We cast the above Hamiltonian in matrix notation: 

\begin{widetext}
\begin{eqnarray} \label{h2}
H_{\mathbf k}^s=  \begin{bmatrix} \hbar C_s (k_z-sQ)+ s \hbar v ( k_z-sQ) & s \hbar \alpha_n(k_x-ik_y)^n \\ 
s \hbar \alpha_n(k_x+ ik_y)^n  & \hbar C_s (k_z-sQ) - s \hbar v ( k_z-sQ) 
\end{bmatrix}.
\end{eqnarray}
\end{widetext}

Hereafter, we use Natural units and set $\hbar=c=k_B=1$. We now examine the effect of circularly polarized light on the mWSM. Under the influence of a periodic optical driving with electric field of frequency $\omega$, ${\mathbf E}(t)= E_0(-\cos\omega t, \sin \omega t, 0) $, the Hamiltonian transforms via the Pierel's substitution $k_i \rightarrow k_i - A_i$, where the vector potential is given by $ {\mathbf A}(t)=\frac{E_0}{\omega} ( \sin \omega t, \cos\omega t,  0)$, in the Landau gauge. The gauge dependent momenta transform as $k_{x} \rightarrow k_{x}^\prime =k_{x} - \: A_0 \sin \omega t$, 
\textcolor{black}{
$k_{y} \rightarrow k_{y}^\prime = k_{y} - \: A_0 \cos \omega t$} ,and  $k_z \rightarrow k_z^\prime =k_z $. The driving amplitude of the vector potential is related to the amplitude of the electric field by $A_0=\frac{E_0}{\omega}$. Considering the fact that $(k'_x \pm ik'_y)^n= \sum_{m=1}^n (k_\perp e^{\pm i\phi})^{n-m} (A_0)^m e^{\pm i m (\frac{\pi}{2}-\omega t )} ~^n C_m$, where $~^n C_m = \frac{n!}{(n-m)! m!}$ 
represents the combinatorial operator, the time dependent Hamiltonian takes the form
  
\begin{eqnarray}
 H_{\mathbf k}^s({\mathbf A},t)&=& s \sigma_+ (k'_x +ik'_y)^n + s \sigma_- (k'_x - ik'_y)^n  \\ \nonumber
 &+&   C( k_z-sQ) + v ( k_z - sQ) \sigma_z 
 \label{h1_time}
\end{eqnarray}

Solving the problem with a time-dependent potential may be out of the reach of analytical tractability. Instead, we resort to using Floquet's theorem and the extracting the sub-leading order term in the high frequency van-Vleck expansion, to obtain a closed form expression for the effective Hamiltonian $H^{F}_{\mathbf k}$.
\textcolor{black}{We note that one can numerically solve an extended 
Floquet Hamiltonian, defined in the Hilbert space ${\mathcal T} \otimes {\mathcal H}$ (with ${\mathcal H}$ being the Hilbert space of static Hamiltonian and ${\mathcal T}$ being the
Hilbert space associated with multi-photon dressed states), to obtain the quasi-states and quasi-energies \cite{shirley68}. From mathematical point of view, one can also use Lie algebra technique and a decomposition of the evolution on each group generator to obtain an effective Hamiltonian \cite{santana19}.
However, the van-Vleck expansion is more tractable as the whole Hilbert
space of extended Floquet Hamiltonian gets projected onto the zero-photon subspace: ${\mathcal T} \otimes {\mathcal H} \to {\mathcal T}_0 \otimes {\mathcal H} = {\mathcal H}$. All the eigenvectors of the effective Hamiltonian, obtained from van-Vleck expansion,
are the projection of the original eigenvectors onto
the model space $ {\mathcal H}$ with the true quasienergies. This shows the real usefulness of  the van-Vleck expansion
 where one can get a closed form expression under high frequency approximation. 
}

 In this  limit, one can describe the dynamics of the driven system over a period $T$ in terms of the effective Floquet Hamiltonian: $H^F_{\mathbf k}
\approx H_{\mathbf k}^s + V_{\mathbf k}^{s}$, where $V_{\mathbf k}^s$ represents perturbative driving term. We restrict to contributions of order $1/ \omega$ throughout the manuscript, and the form of $V_{\mathbf k}^s$ is given by  
\begin{equation}
\label{h4}
V^s_{\mathbf k}=\sum_{p=1}^\infty \frac{[V_{-p},V_p] }{p \omega},
\end{equation}

with $V_p=\frac{1}{T} \int_0^T H_{\mathbf k}^s({\mathbf A},t) \: e^{i p \omega t} \: dt $ and $\omega=\frac{2 \pi}{T}$. Evaluating $V^s_{\mathbf k}$ for our system, we arrive at

\begin{eqnarray} \label{v1}
V_p &=& s\alpha_n \sum_{m=1}^n  (k_\perp)^{n-m}(-A_0)^m ~^n C_m \nonumber \\ 
&=&  \begin{bmatrix} 0 & e^{-i [(n-m) \phi +m\frac{\pi}{2}]} \delta_{p,-m} \\  
  e^{i [(n-m) \phi +m\frac{\pi}{2}]} \delta_{p,-m} & 0 \end{bmatrix}.
\end{eqnarray}

Using the result in (\ref{v1}) and evaluating the commuatator in (\ref{h4}), we find that the effective Floquet Hamiltonian takes the form
\ba \label{h5}
H^F_{\mathbf k} & = & H_{\mathbf k}^s+ V_{\mathbf k}^s \nonumber \\ 
&& = C_s(k_z-s Q) + s \alpha_n {\boldsymbol{\sigma}} \cdot ( {\bf n}_{\mathbf k}- s Q {\hat{e}}_z) \nonumber  \\ 
& +& \frac{\alpha_n^2}{\omega} \sum_{p=1}^n \frac{1}{p} \bigg( ^nC_p A_0^p\bigg)^2 k_\perp^{2n-2p }\sigma_z  \nonumber \\
&=&  C_s(k_z-s Q) + s \alpha_n (\mathbf{n}'_{\mathbf {k}}-s Q {\hat{e}}_z) \cdot {\boldsymbol \sigma} 
\ea
with $\mathbf{n}'_{\mathbf {k}}=\left(  k^n_{\bot}\cos \left( n \phi_{k} \right), k^n_{\bot}\sin 
\left( n \phi_{k} \right), T_{\mathbf k}/\alpha_n \right)$. 
In all susequent analysis, we define $T_{\mathbf k} \equiv v k_z + \frac{\alpha_n^2}{\omega} \sum_{p=1}^n \beta_p^n k_\perp^{2(n-p)} \equiv \Delta_n + 
T'_{\mathbf k}$, with $  T'_{\mathbf k} \equiv v k_z + \frac{\alpha_n^2}{\omega} \sum_{p=1}^{n-1} \beta_p^n k_\perp^{2(n-p)} $,
and $\beta_p^n=(^n C_p A_0^p)^2/p$. The momentum independent contribution to the Floquet Hamiltonian acquires
the form $\Delta_n= \frac{\alpha_n^2 A_0^{2n}}{n \omega} $. It clear from the construction of (\ref{h5}) that the effective Hamiltonian embodies
terms which couple higher momentum modes (modes which diverge faster than $k$ as $k \rightarrow \infty$) of the Weyl fermion to the photon.
{ This can induce a ``{non-renormalizable}" nature to the theory, which as we shall see, becomes strongly dependent on a momentum cutoff.
This phenomenon is very similar to the theory of quantum electrodynamics
with massive operators in high energy physics.} The extra terms, absent for $n=1$, appears due to the 
anisotropic energy dispersion of the static mWSM Hamiltonian (\ref{h1}). A close inspection of effective Hamiltonian 
(\ref{h5}) suggests that circularly polarized light can not open up a gap in WSM as the time reversal symmetry is 
intrinsically broken in static Hamiltonian (\ref{h1}); instead the position of the Weyl points shifts from $(0,0,sQ) \to 
(0,0,sQ -\Delta_n)$. We note here $T'_{\mathbf k}=v k_z$ for ${\mathbf k}=(0,0,k_z)$. Interestingly, 
unlike the single Weyl case where  the shift $Q$
 quadratically varies with driving amplitude $A^2_0$, the shift in the Weyl point for
 mWSMs $\Delta_n$ is coupled with 
monopole charge $n$ as $\Delta_n= \frac{\alpha_n^2 A_0^{2n}}{n \omega} $. \textcolor{black}{Therefore, Weyl points receive a topological charge 
dependent shift under irradiation.}
The terms containing $k_{\perp}$ in 
$T'_{\mathbf k}$ would lead to subleading corrections in transport properties.\\

The effective quasi-energies obtained from effective Floquet Hamiltonian (\ref{h5}), are thus 
\ba
E^F_{{\mathbf k}} = C_s(k_z-s Q) \pm s \sqrt{ \alpha^2_{n} k^{2 n}_{\bot} + T^2_{\mathbf k}},
\label{eff_eng}
\ea
leading to the established result: $ E^F_{{\mathbf k}}(n=1)=C_s(k_z-s Q) \pm s \sqrt{ v^2 k^{2 }_{\bot} 
+ (vk_z + \Delta_1)^2}$ as $\alpha_1=v$.
One can observe that $k_{\perp}$ term  in $T_{\mathbf k}$ is absent for conical dispersion while for $n>1$, the distortion anisotropy
in conical dispersion leads to terms dependent on  $k_{\perp}$  in $T_{\mathbf k}$. 
For completeness, we note that the static energy of an mWSM Hamiltonian with no
driving is obtained by diagonalizing the Hamiltonian (\ref{h1}) to obtain
$E^0_{\mathbf k}= C_s(k_z-s Q) \pm \sqrt{\alpha^2_n k^{2n}_{\bot} + v^2 k^2_z}$.
Therefore, one can clearly see that the external optical field paramters get coupled with 
momentum $k_{\perp}$ leading to the complicated form of $T_{\mathbf k}$ in Eq.~(\ref{eff_eng}).
\textcolor{black}{ In particular,  the nature of the Floquet dispersion (\ref{eff_eng}) changes due to the coupling of the incident light parameter $A_0$ and $\omega$ with the momentum 
$k_{\perp}$ and the topological charge $n$. Another interesting 
feature of the Floquet dispersion is that $k_z$ gets coupled to $k_{\perp}$ that is not noticed for irradiated single WSMs. 
Apart from these characteristic changes, Floquet spectrum remains gapless at $(0,0,\pm Q-\Delta_n)$.
The extensive analysis of Floquet dispersion can be shown to exhibit a few distinct behavior as compared to static dispersion  
\cite{supple}.}\\


\section{  Berry Curvature}
\label{Berry_curvature}

It is very important to study geometric phases in any topological system as the 
anomalous response function is directly given by the Berry curvature. Here our aim would be 
to investigate the effect of the driving on the Berry curvature and subsequently on the anomalous 
transport. Before going into detail,
we begin by defining the Berry curvature associated with the 
Floquet Hamiltonian $H^F_{\mathbf k}$. 
The Berry curvature of the m$^{\textrm{th}}$ band for a Bloch Hamiltonian $H(k)$,
defined as the Berry phase per
unit area in the $k$ space, is given by ~\cite{Xiao_06}
\begin{equation}
\Omega^{m}_{a} (\mathbf{k})= (-1)^m \frac{1}{4|n_{\mathbf{k}}|^3} \epsilon_{a b c} \mathbf{n}_{\mathbf{k}} 
\cdot \left( \frac{\partial \mathbf{n}_{\mathbf{k}}}{\partial k_b} \times \frac{\partial \mathbf{n}_{\mathbf{k}}}{\partial k_c} \right) .
\label{bc}
\end{equation}

The explicit form of the Berry curvature associated with the Weyl node
{\textcolor{black}{having chirality $s$}} as obtained from Floquet effective Hamiltonian 
(\ref{h5}) is given by
\begin{eqnarray}
\textcolor{black}{
{\Omega}^{\pm,s}_F({\mathbf{k}})} &=&\pm \frac{1}{2} \frac{1}{|E^F_{{\mathbf k}}|^3}
( n v \alpha_n^2 k^{2n-1}_{\bot} \cos \phi_{\mathbf k}, 
 n v \alpha_n^2 k^{2n-1}_{\bot} \sin \phi_{\mathbf k}, \nonumber \\ 
&-&n \beta_{\mathbf k} \alpha_n^2 k^{2n}_{\bot}  +
 T_{\mathbf k} n^2 \alpha_n^2 k^{2n-2}_{\bot}  
 ),
\label{eq_bcl}
\end{eqnarray}
with $\beta_{\mathbf k}= \frac{\alpha_n^2}{\omega} \sum_{p=1}^n (2n-2p) \beta^n_p k_\perp^{2n-2p-2 }$. 
\textcolor{black}{We note that $+(-)$ sign refers to the valence (conduction) band and chirality $s=\pm 1$. The Berry curvature 
remains unaltered irrespective of the chirality of the Weyl nodes i.e., ${\Omega}^{\pm,+}={\Omega}^{\pm,-}$. This is due to the fact that the 
chirality factor $s$ appears in the Hamiltonian  with all $\sigma_i$'s; a close inspection suggests that 
$s$ gets cancelled from numerator and denominator in 
Eq.~(\ref{bc}).}

One can obtain regular static Berry curvature when $A_0=0$,
$\beta_{\mathbf k}=0$ and $T_{\mathbf k}=v k_z$. The static Berry curvature using Hamiltonian (\ref{h1}) becomes
\begin{eqnarray}
\textcolor{black}{
{\Omega}^{\pm,s}_0({\mathbf{k}})} &=&\pm \frac{1}{2} \frac{1}{|E^0_{{\mathbf k}}|^3}
( n v \alpha_n^2 k^{2n-1}_{\bot} \cos \phi_{\mathbf k}, 
 n v \alpha_n^2 k^{2n-1}_{\bot} \sin \phi_{\mathbf k}, \nonumber \\ 
&& n^2 v \alpha_n^2 k^{2n-2}_{\bot} k_z
 ),
\label{eq_bcl}
\end{eqnarray}
Therefore, one can observe that ${\Omega}_z({\mathbf{k}})$ is modified 
due to the driving, while the remaining two components of ${\Omega}^{\pm}_F({\mathbf{k}}) $
receives the correction from the effective  energy $E^F_{{\mathbf k}}$ appearing in the denominator. 
This suggests that anomalous conductivity $\sigma^a_{xy}$ would be heavily modified due to the driving 
as compared to $\sigma^a_{xz}$ and $\sigma^a_{yz}$. We shall analyze this extensively in the what follows.

Now, turning to $n=1$ case,
the Berry curvature for driven single WSM case is given by 
${\Omega}^{{\pm},s}_F({\mathbf{k}},n=1)=(k_x,k_y,k_z+ v^3 \Delta_1)/|E^0_{{\mathbf k}}(n=1)|^3$.
One can clearly observe that for driven mWSMs  all components of ${\Omega}({\mathbf{k}})$ depend on 
$k_{\perp}$, while for  driven single WSM case individual components are comprised of separate momentum.
Thus the dispersion anisotropy of the $n>1$ mWSMs imprints effects which are absent for the single WSM case. 
Importantly, even for ${\Omega}_z({\mathbf{k}})$ in single WSMs, the momentum independent term 
$\Delta_1 \sim A_0^{2}$ bears the signature of periodic driving.  For $n>1$, the topological charge gets 
coupled with the driving paramter which leads to a more complex form of ${\Omega}_z({\mathbf{k}})$ as compared to the $n=1$ case.

{We shall compute the anomalous Hall conductivity $\sigma^a_{F,xy}$, considering the effective 
Floquet Hamiltonian, from the $z$-component of 
Berry curvature in Eq.~(\ref{eq_bcl}). In order to obtain a closed form results in the  leading order, 
we neglect $\beta_{\bm k}$ as $\omega \to \infty$ as the effective energy in the 
denominator bears the correction terms due to driving as shown in Eq.~(\ref{eff_eng}).
We, on the other hand, consider the effect of the Floquet 
driving on the cut-off limit of $k_z$ integration. 
In particular, $z_l= -\Lambda- sQ \to z'_l$
and $z_u= \Lambda- sQ \to z'_u$ with  $z'_l=-\Lambda-sQ+s \Delta_n$ and 
$z'_u=\Lambda-sQ+s \Delta_n$. Therefore, one can 
safely consider the static energy in the denominator, and we shall motivate this assumption extensively while discussing the 
vacuum contribution Sec.~\ref{vc}. The anomalous contribution to leading order
is thus given by
\begin{eqnarray}
 \sigma^a_{F,xy}&=& e^2\int \frac{d {\bm k}}{4\pi^2} \sum_s  \textcolor{black}{\Omega^{-,s}_F ({\bm k})} \nonumber \\
 &\simeq& - \frac{n  e^2}{4 \pi^2} \int^{z'_u}_{z'_l}\int^{\infty}_0 dk_{\perp} dk_z
 \frac{k_z   k_{\perp}}{(k_z^2 + k^2_{\perp})^{3/2}} \nonumber \\
 &\simeq& -\frac{n  e^2}{2 \pi^2}(Q + \Delta_n) 
 \label{eq_anomalous}
\end{eqnarray}
We have considered cylindrical polar co-ordinates for the convenience of the integration along with the following rescaling: $k_z\to k_z/v$ and $k_\perp \to k^{1/n}_{\perp} \alpha^{-1/n}_n$.
It is noteworthy that this anomalous Hall coefficient has a topological property due to the appearance of the monopole charge. For the static system, it is just given by $-\frac{n  e^2}{2 \pi^2} Q$. 
} 
\textcolor{black}{ Since the Berry curvature of the filled valence band remains 
same for both the nodes with opposite chiralities.
The results obtained considering these two nodes is just the double of that of the obtained in single node.}
\\

{\color{black} We now connect our findings to the transport phenomena in the mWSMs. It has been shown that there exist $n$ number of  Fermi arcs for a mWSM with topological charge   
$n$ \cite{rmdantas19}, and we know the transport is mainly governed by the surface states present in the Fermi arc for WSMs}. Interestingly, driving shifts the position of Weyl points $\pm Q\to \pm Q  + \Delta_n$; this leads to the modification in Fermi arc for irradiated case as compared to the static case. As a result, transport properties receive additional corrections from driving. It has been shown that Fermi arc can be tuned using  Floquet replica technique when a WSM is irradiated with circularly polarized light \cite{oka17}. The factor $n$ in front of Eq.~(\ref{eq_anomalous}) signifies that effective Floquet Hamiltonian still supports $n$ number of Fermi arcs. We here mention 
that the neglected  $\beta_{\bm k}$ term would give rise to sub-leading non-topological 
contributions. Since we wish to probe the question of transport due to laser driving, 
it would be appropriate to investigate the optical conductivity using Floquet-Kubo formalism. 
However,  we note at the outset that  one can find similar expression as given in Eq.~(\ref{eq_anomalous}) while calculating the  vacuum contribution of optical 
conductivity up to leading order.

\section{Conductivity tensor} \label{sigma_xy}



Having derived the Berry curvature induced anomalous Hall conductivity, we shall now systematically 
formulate the conductivity tensor using the current-current correlation function. This is 
constructed using the Matsubara Green’s function method. 
The current-current correlation is written  as
\ba
\prod_{\mu \nu}(\Omega, {\bf k})&=& T \sum_{\omega_n} \sum_{s=\pm} \int \frac{d^3k}{(2\pi)^3}
J_\mu^{(s)} G_s(i \omega_n, {\bf k}) \nonumber \\
&&J_\nu^{(s)} G_s (i \omega_n-i\Omega_m, {\bf k}-{\bf q}) |_{i \Omega_m \rightarrow \Omega+i \delta}
\label{c-c}
\ea
Here, $ \mu, \nu =\{x,y,z\}$, $T$ is the temperature, $\omega_n$ and $\Omega_n$ are 
the fermionic and bosonic Matsubara frequencies and $G$
is the single particle Green's function. The Hall conductivity
can  now be derived from the zero frequency $\Omega \to 0$ and zero wave-vector limit. 
\\

{  Using the current-current correlation (\ref{c-c}), one can define the 
static  conductivity tensor $\sigma^0_{ab}$.  
We here use the form of the time-averaged conductivity tensor $\sigma^F_{ab}$ in the form of the Kubo formula, modified for the Floquet states  as
\begin{eqnarray} \label{cond1}
\sigma^F_{ab} =& i& \int \frac{d^3{\bf k}}{(2\pi)^3} \sum_{\alpha \neq \beta} \frac{f_\beta({\bf k}) - f_\alpha ({\bf k})}{\epsilon_\beta ({\bf k}) - \epsilon_\alpha ({\bf k})}  \nonumber \\
&\times& \frac{\langle {\Phi_\alpha  ({\bf k}) | J_b | \Phi_\beta  ({\bf k}) } \rangle \langle {\Phi_\beta  ({\bf k}) | J_a | \Phi_\alpha  ({\bf k}) } \rangle}{\epsilon_\beta ({\bf k}) - \epsilon_\alpha ({\bf k}) + i\eta }
\end{eqnarray}
which resembles the standard form of the Kubo
formula  where $J_{a(b)}$  represents  the  current  operator,
the $|\Phi_{\alpha}({\mathbf k})\rangle$  represents  the  states  of  the  effective  Floquet
Hamiltonian (\ref{h5}),  and   $\epsilon_{\alpha}$  represent  the  corresponding
quasi-energies. The $f_{\alpha}$  represent   the   occupations
which  in  general  could  be  non-universal  in  systems
which are out of equilibrium.  In such cases, the steady-state
occupations can take the form of Fermi-Dirac distribution associated with the
quasi-energies of the Floquet states, depending on the characteristics of the drive. The Matusubara formalism turns out to hold for Floquet states as well \cite{menon18}. 
\textcolor{black}{The method of Floquet Kubo formalism has been widely used in
calculating optical Hall conductivity in open and closed quantum systems.
\cite{floquet_kubo}}. \\

One can start from
Luttinger’s phenomenological transport equations \cite{luttinger64} for the
electric and energy DC currents. 
The energy  current is  originated  from  the
combination of heat current $J_Q$ and energy transported by 
the electric current $J_E$ in presence of 
 electromagnetic field while the underlying system is characterized by a finite
 chemical potential $\mu$ and temperature $T$. 
 Within the  Fermi  liquid  limit $k_B T \ll |\mu|$, 
 the Mott rule and the Wiedemann-Franz law relate the
 thermopower $\alpha$ and thermal conductivity $K$, respectively,
 to the electric conductivity $\sigma$
 \cite{ferreiros17,lundgren14,jonson80}:
 \begin{equation}
 \alpha_{ab}= e L T \frac{d \sigma_{ab}}{d \mu},~~~~ K_{ab}= L T \sigma_{ab}.
  \label{formula}
 \end{equation}
 Here, $\alpha_{ab}$ is the Nernst conductivity and $K_{ab}$ is the
 thermal Hall conductivity and 
 $L=\pi^2 k_B^2/3 e^2$ is the Lorentz number.
These formulas are assumed to be valid for the effective time-independent Floquet Hamiltonian setup \cite{menon18}, and we shall investigate them in what follows. \\

One can define the current operator from the effective Floquet Hamiltonian 
$H^F_{\mathbf k} $  Eq.~(\ref{h5})
\be \label{j1}
J_\mu=e \frac{\partial H^F_{\mathbf k}}{\partial k_\mu}
\ee
In order to derive $J_\mu$, we consider the leading order term 
neglecting ${\partial T_{\mathbf k}}/{\partial k_\mu}$ term as it contains
$1/\omega$ factor. We note that the current operator obtained in this manner would be the same as the static current operator for 
mWSM Hamiltonian. This leading order term can be further
confirmed by the zeroth order Fourier component of the current operator as shown in the 
SI, Sec. III. 
However, one can indeed consider the full current operator
with ${\partial T_{\mathbf k}}/{\partial k_\mu}$ to obtain the higher order 
corrections. The effect of $T_{\mathbf k}$ term
is also encoded in the single particle
Greens function $ G_s$.
We compute the optical conductivity by using the complete expression of  $G$
and approximated  current operator.\\

In terms of $\sigma$'s, we can write upto leading order as 
\be \label{jx3}
J_x \approx e s n \alpha_n  k_\perp^{n-1}[ \cos((n-1)\phi_{\mathbf k})\sigma_x +\sin((n-1)\phi_{\mathbf k})\sigma_y   ]
\ee
\be \label{jy3}
J_y \approx e s n \alpha_n k_\perp^{n-1}[ \cos((n-1)\phi_{\mathbf k})\sigma_y -\sin((n-1)\phi_{\mathbf k})\sigma_x ]
\ee
The point to note here is that $J_{x}$ and $J_y$ both depend on $k_x$ and $k_y$ which is in contrast to the single WSM case where 
$J_i \sim k_i \sigma_i$. The anisotropic nature of dispersion of the 
mWSM Hamiltonian thus engravs its effect on the current operator. \\

Employing the current-current correlation and performing a detailed calculation
\cite{supple}, we arrive at the conductivity tensor as
\ba
\sigma_{x y}&=& \frac{e^2 n^2 \alpha_n^2}{4 \pi^2} \sum_{s=\pm} \int_0^\infty dk_\perp k_\perp^{2n-1} \int_{-\Lambda}^{\Lambda} dk_z \nonumber \\
&&
\frac{s v (k_z-Q)+\frac{s \alpha_n^2}{\omega} \sum_{p=1}^n \beta^n_p k_\perp^{2(n-p)}}{[(\frac{s\alpha_n^2}{\omega} \sum_{p=1}^n \beta^n_p k_\perp^{2(n-p)}
+ s v (k_z-sQ))^2+ \alpha_n^2 k_\perp^{2n}]^{3/2}} \nonumber \\
&\times& 
[n_F(E_{\mathbf k}^{F,-}) -n_F(E_{\mathbf k}^{F,+})]
\label{eq_oc_total}
\ea
{where $\Lambda$ is the ultra-violet
cut-off of $k_z$ integral},
$n_F(E)=\frac{1}{e^{\beta(E-\mu)}+1}$ is the Fermi-Dirac distribution function, and $\beta = 1/T$ is inverse temperature. The total 
optical conductivity (\ref{eq_oc_total}) is the sum of vacuum and Fermi surface contributions that we shall extensively calculate below.
\textcolor{black}{We note that due to the existence of external and internal energy scale $\omega$ and $\mu$, the cut-off $\Lambda$ plays an 
important role in achieving physically meaningful results.
This cut-off is ultra-violet in nature and can  in principle depend on the detail of the material.}

\subsection{Vacuum contribution} \label{vc}

In this section, we  investigate 
the vacuum contribution which is obtained in the limit 
$[n_F(E_{\mathbf k}^{F,-})-n_F(E_{\mathbf k}^{F,+})] \rightarrow 1 $. 
Physically this means that valence (conduction) band is completely filled
(empty). This vacuum contribution amounts for an
intrinsic contribution that remains finite for $\mu \to 0$.
Computationally, this refers to the situation where the upper
limit  $k_{\perp}$ is considered to be $\infty$ in the literature. 
 With suitable redefinitions and linear integration variable shifts, we arrive at 

\ba
\sigma_{xy}^{vac} &=& \frac{e^2 n^2 \alpha_n^2}{4 \pi^2} \sum_{s=\pm} \int_0^\infty dk_\perp k_\perp^{2n-1} \int_{-\Lambda-sQ+\frac{s \alpha_n^2}{\omega}A_0^{2n}}^
{\Lambda-sQ+ \frac{s \alpha_n^2}{\omega}A_0^{2n}} \nonumber \\
&\times& \frac{[s v k_z + \frac{\alpha_n^2}{\omega}\sum_{p=1}^{n-1} \beta^n_p k_\perp^{2(n-q)}]}
{[(\frac{\alpha_n^2}{\omega}\sum_{q=1}^{n-1} \beta^n_q k_\perp^{2(n-q)})^2+ \alpha_n^2 k_\perp^{2n}]^{3/2}}.
\ea

We will compute the vacuum contribution using two separate procedures involving suitable approximations and then compare the results obtained. 

\subsubsection{Coordinate Transformation Method}

The method prescribed in this section relies on the fact that 
while several quantities are set to infinity in a computation, in order to get physically plausible answers one might need to define the order in which the limits are taken. For computation of the integrals, the following coordinate map $\mathcal{M} : \mathbb{R}^2 \rightarrow \mathbb{R}^2$ is prescribed with the action $k_{\perp } \to k'_{\perp} =  k^{\frac{1}{n}}_{\perp} \alpha^{-\frac{1}{n}}_n$, and $k_z \rightarrow k_z$. With this coordinate transformation, the vacuum contribution of the conductivity tensor looks like
 
\ba \label{vac1}
\sigma_{xy}^{vac}= - \frac{e^2 n \alpha^{2-\frac{2}{n}}_n}{4 \pi^2} 
\sum_{s=\pm}s  \int_{z_l}^{z_u}  \int_{x_l}^{x_u} 
\frac{k_{\perp} T_{\mathbf k} }{(k_{\perp}^2+ T_{\mathbf k}^2)^{3/2}}dk_{\perp} 
 dk_z \nonumber \\
\ea
Here, the upper and lower limits of the integrals have been determined
with appropriate physical justifications \cite{supple}:
\ba \label{limits}
x_l & = & 0, ~~~~ x_u  =  \Lambda_\perp  \\ \nonumber
z_u & = & v( \Lambda - sQ )+ s \left( \Delta_n + \frac{\alpha_n^2}{\omega} \sum_{p=1}^{n-1}\beta_p^n 
 \alpha^{\frac{2(p-n)}{n} }_n 
\Lambda^{\frac{2(n-p) }{n}}_{\perp} \right) \\ \nonumber
z_l & = &  v( -\Lambda- sQ )+ s \left(\Delta_n +  \frac{\alpha_n^2}{\omega} \sum_{p=1}^{n-1}\beta_p^n 
\alpha^{\frac{2(p-n)}{n} }_n  
\Lambda^{\frac{2(n-p) }{n}}_{\perp} \right)
\ea
$\Lambda_{\perp}$ is the cut-off for $k_{\perp}$ integral. 
One can segregate $z_{l,u}$ from $\Lambda_{\perp}$: $z_{l,u}=z'_{l,u} + s \frac{\alpha^2_n}{\omega} X_{\Lambda_{\perp}}$
with $X_{\Lambda_{\perp}}=\sum_{p=1}^{n-1} \beta^n_p \alpha^{\frac{2(p-n)}{n} }_n 
 \Lambda^{\frac{2(n-p) }{n}}_{\perp}$ and $z'_{l,u}=v( \mp \Lambda - sQ )+ s \Delta_n $.
 Hence one has to handle this cut-off with care, and the issue reduces to the order of taking limits.    
\textcolor{black}{ We again stress that the high frequency Floquet effective Hamiltonian is valid when $\omega$ is  larger than the bandwidth not permitting any real electronic transitions.
Keeping this in mind,  the sub-leading $1/\omega$ order correction that we want to extract is preserved as we execute the $k_{\perp}$ integral
followed by the $k_z$ integral.
We note that while solving the $k_{\perp}$ integral, without loss of generality $\Lambda_{\perp}$
is considered to be large as compared to $\Lambda$. Importantly,  $\Lambda_{\perp}/\omega$ is small compared to $\Lambda$ and hence $X_{\Lambda_{\perp}}$ is a sub-leading term since $\omega$ sets the dominant energy scale in the problem. 
Taken collectively, the subleading $X_{\Lambda_{\perp}}$ term is held finite during the $k_{\perp}$ integration and this leads to the $\Lambda_{\perp}$ dependence reappearing through the limits of the $k_z$ integral.  In a nutshell, our resullt is applicable when $\omega \gg \Lambda_{\perp}  \gg \Lambda $.
We justify the above assumptions for the high frequency 
Floquet effective Hamiltonian $H^F_{\bm k}$ (\ref{h5}) that is derived from a low energy minimal model (\ref{h1}).}  \\

Finally, we obtain the vacuum contribution of conductivity in mWSM,
\be \label{vacf}
\sigma_{xy}^{vac}= n \frac{e^2 Q \alpha^{2-\frac{2}{n} }_n }{2 \pi^2} -
n \frac{e^2 \alpha^{2-\frac{2}{n} }_n }{2 \pi^2}\Bigg[\Delta_n - \frac{\alpha_n^2}{\omega v}
\sum_{p=1}^{n-1} \beta_p^n \Lambda_\perp ^{2(n-p)}\Bigg]
\ee
Here, $\Delta_n$ and $\beta_p^n$ are the contributions appearing as an effect of light.
\textcolor{black}{
For mWSMs, the light induced
Weyl node position depends on the topological
charge associated with the Weyl node. This shift in Weyl nodes 
reduces to a driving paramter dependent constant value  
as observed in the irradiated single WSMs}. One can easily recover the $n=1$ 
behavior of gap where $\Delta_1$ varies quadratically with 
the amplitude of driving $A_0$ \cite{menon18}. For $n> 1$ further corrections, due to higher order curvature of the Floquet Hamiltonian, contribute in terms of the cut-off of the low-energy model. 

\subsubsection{Series Expansion Method}

We shall now proceed with a physically justified  alternative method to compute $\Lambda_{\perp}$ in terms of the $k_z$ cut-off. The idea here is 
to expand the denominator around its unperturbed static energy in increasing 
powers of driving period $1/\omega \to 0$ as $\omega \to \infty$. 
The  perturbative expansion is then given by
\begin{equation}
 k^2_{\perp} +T_{\mathbf k}^2 \approx E^2_{\mathbf k} + \frac{2 v k_z \alpha^2_n}
 {\omega} \sum_{p=1}^{n-1}\beta^p_n  
 \alpha^{\frac{2(p-n)}{n} }_n 
 k^{\frac{2(n-p) }{n}}_{\perp}
 \label{eq_alt1}
\end{equation}
One can then note that for $n=2$, only $\beta^n_1$ exists while for $n=3$, 
$\beta^n_1$ and  $\beta^n_2$ both exist.
$E_{\mathbf k}=\sqrt{k^2_{\perp} + v^2 k_z^2}$ is the bare static energy
of single WSM in the absence of tilt.  
Considering $X_{k_{\perp}}=\sum_{p=1}^{n-1} \beta^n_p \alpha^{\frac{2(p-n)}{n} }_n 
 k^{\frac{2(n-p) }{n}}_{\perp}$,
we now express the integrand as  
\begin{eqnarray}
 \frac{T_{\mathbf k}}{(k^2_{\perp} + T^2_{\mathbf k} )^{3/2}} 
 &\approx& \frac{1}{E^3_{\mathbf k} } \Bigg(vk_z -
 \frac{3 v^2 k_z^2 \alpha^2_n}{E^2_{\mathbf k}\omega } X_{k_{\perp}}\nonumber \\
 &+& \frac{\alpha^2_n}{\omega}  X_{k_{\perp}} 
 \Big(1- \frac{3 v k_z \alpha^2_n}{E^2_{\mathbf k}\omega } X_{k_{\perp}}\Big)
 \Bigg)
 \label{eq_alt2}
\end{eqnarray}

We explicitly write $\sigma_{xy}^{vac}$ for $n=2$
(neglecting $1/\omega^2$ term)  as,
\begin{eqnarray}
  \label{vac1}
&&\sigma_{xy}^{vac}(n=2)= - \frac{e^2 n \alpha^{2-\frac{2}{n}}_n}{4 \pi^2} 
\sum_{s=\pm}s  \int_{z'_l}^{z'_u}  \int_{0}^{\infty} \nonumber \\
&&
\frac{k_{\perp} T_{\mathbf k} }{(k_{\perp}^2+ T_{\mathbf k}^2)^{3/2}}dk_{\perp} 
 dk_z  \nonumber \\
 &\approx& -\frac{e^2 n \alpha^{2-\frac{2}{n}}_n}{4 \pi^2} \sum_{s=\pm}
 s\Bigg(v (z_l +z_u) + {v^2 \alpha_n \beta^n_1} (z'_l-z'_u) \Bigg) \nonumber \\
 &\approx& -\frac{e^2 n \alpha^{2-\frac{2}{n}}_n}{4 \pi^2}  
 \Bigg(v(-2  Q + 2  \Delta_n) + 
 \frac{2  v^2 \alpha_n \beta^n_1}{\omega} \Lambda\Bigg) 
 \label{eq_alt3}
\end{eqnarray}
In this derivation, we ignore the divergent contributions 
coming from the integrals having higher powers of $k_{\perp}$ in the numerator. 
These types of terms, being artifacts of the 
underlying low-energy model, do not appear in the lattice model. In order to 
obtain $\Lambda_\perp$, we equate the coefficient of $1/\omega$ from 
Eq.~(\ref{eq_alt3}) and Eq.~(\ref{vacf}).  We find $\Lambda_{\perp}$
linearly depends on $\Lambda'$: $\Lambda_{\perp}= 2 v^2 \Lambda'$. 
For $n=3$, we find
\begin{eqnarray}
&&\sigma_{xy}^{vac}(n=3) = - \frac{e^2 n \alpha^{2-\frac{2}{n}}_n}{4 \pi^2} 
\Bigg[v(-2  Q + 2 \Delta_n) \nonumber \\
&-&\frac{2 v^2 \alpha^{\frac{2}{n} }_n \beta^n_1 }{\omega \sqrt{\pi}}
\Gamma\Big(\frac{5}{6}\Big) \Gamma\Big(\frac{5}{3}\Big) 
\Big(\frac{|z_l|^{\frac{4}{n}} + z_u^{\frac{4}{n}} }{4/3} \Big)\nonumber \\
&+& \frac{\beta^n_2}{\omega} \Bigg[ -\frac{3 v^2 \alpha^{\frac{4}{n} }_n}{2 \sqrt{\pi} }
\Gamma\Big(\frac{7}{6}\Big) \Gamma\Big(\frac{7}{3}\Big)  
\Big(\frac{|z_l|^{\frac{2}{n}} + z_u^{\frac{2}{n}} }{2/3} \Big) \nonumber \\
&+& \frac{\alpha^{\frac{4}{n} }_n}{\sqrt{\pi} } 
\Gamma\Big(\frac{1}{6}\Big) \Gamma\Big(\frac{4}{3}\Big)
  \Big(\frac{|z_l|^{\frac{2}{n}} + z_u^{\frac{2}{n}} }{2/3} \Big)  \Bigg]
\Bigg]
\label{eq_alt4}
\end{eqnarray}
 It is noted that contrary to the  $n=2$ case,  $\Lambda_{\perp}$ is 
non-linearly related to $\Lambda'$ for $n=3$: $\Lambda_{\perp}^{\frac{4}{n}}=
\eta_1[  |z_l|^{\frac{4}{n}} + z_u^{\frac{4}{n}}]$ and 
$\Lambda_{\perp}^{\frac{2}{n}}=
\eta_2 [ |z_l|^{\frac{2}{n}} + z_u^{\frac{2}{n}}]$ where $\eta_{1,2}$
can be obtained by matching the coefficient of $\beta^n_1/\omega$ and 
$\beta^n_2/\omega$. \textcolor{black}{The relationship between $\Lambda$ and $\Lambda_{\perp}$ derived here are consistency conditions for the model parameters.}

\subsection{Fermi-surface contribution}
\label{fs_contribution}

We take a note of the point that for the calculation of the Fermi surface contribution, one has to consider the finite upper limit in the   $k_{\perp}$ integral as $b$, a parameter which we compute below. The Fermi surface contribution 
for a given $n$ becomes
\begin{eqnarray}
\sigma_{xy}^{FS}(n) &=&  n \alpha^{2-2/n}_n \sum_s s   \int_{ z'_l}^{z'_u} dk_z \int_0^b \frac{k_\perp T_{\mathbf k} }
{(k_\perp^2 + T^2_{\mathbf k} )^{3/2}} dk_\perp \nonumber \\
&\times& \{\Theta( v^2k_z^2 + (C k_z + s C \Delta_n- \mu)^2)-1 \}
\end{eqnarray}
In the equation above, $\Theta(x)$ represents the Heaviside function which arises from the zero-temperature Fermi-Dirac distribution. It is then more convenient to write $T_{\mathbf k}$ explicitly for 
 $n=2$ as   $T_{\mathbf k}= v k_z  +  \beta^n_1  \alpha^{-2/n}_n
 k_\perp^{2/n}$ and for $n=3$ as
 $T_{\mathbf k}= v k_z  +  \beta^n_1 \alpha^{-4/n}_n
 k_\perp^{4/n} +  \beta_2^n \alpha^{-2/n}_n k_\perp^{2/n}$.
In a more compact notation,  for $n=3$, we define
$ \beta_2^\prime= \beta^n_2 
\alpha^{-2/n} $,  $\beta_3^\prime= \beta^n_1 \alpha_n^{-4/n}$ and 
for $n=2$,  we define $ \beta_2^\prime= \beta^n_1
\alpha^{-2/n} $ and $\beta_3^\prime=0$. 
On the other hand, $b= \{ C k_z + s C \beta_1- \mu)^2 -v^2 k_z^2 \}^{1/2} $. 
Below we shall exprees all our findings in terms of 
$\beta'_2$ and $\beta'_3$ for a general $n$.\\

In the leading order approximation, $\beta_1={\mathcal O}
(\frac{1}{\omega})$, $ k_z \rightarrow 0$ and $\mu$ is held finite.
We shall consider the cases for type-I and type-II cases separately: 
$|C| \gg v, b=\mu-Ck_z $ and $|C|\ll v, b= (\mu^2-v^2 k_z^2 )^{1/2}$.
We again make resort to leading order method where we permit 
${\mathcal O}(1/\omega)$ order term and obtain the following: 
\begin{equation}
\label{condfs}
\sigma_{xy}(n)= \frac{n \alpha^{2-2/n}_n}{v} \int_{z'_l}^{z'_l} d k_z \int_0^b  dk_\perp  k_\perp ( F_{{\mathbf k,1}} +
F_{{\mathbf k},2}+F_{{\mathbf k},3} )
\end{equation}
with 
$
 F_{{\mathbf k},1} = \frac{k_z}{E^3_{\mathbf k} }, F_{{\mathbf k},2} = \frac{\beta'_2 k^{\frac{2}{n} }_{\perp}
+ \beta'_3 k^{\frac{4}{n} }}{E^3_{\mathbf k} },
F_{{\mathbf k},3} = -\frac{3 k_z^2 F_{{\mathbf k},2}}
{E^5_{\mathbf k}}.
$
We note that in Eq.~(\ref{condfs}) the leading order term $F_{{\mathbf k},1}$ is 
also present for the $n=1$ Weyl node case. Similar to the vacuum contribution of optical 
conductivity, 
the multi Weyl nature appears here through a multiplicative 
factor $n \alpha^{2-2/n}_n$. The additional anisotropic and band bending
corrections appear in terms of $1/\omega$ in
$F_{{\mathbf k},2}$ and $F_{{\mathbf k},3}$. {
To obtain a minimal expression,  the above derivation is simplified by neglecting the term  $F_{{\mathbf k},3}$ as 
$k_z^2/{E^3_{\mathbf k}} \to 0$ for $k_z \to 0$ considered for low-energy model. }\textcolor{black}{A close inspection suggests that $F_{k,3}$ contains ${\mathcal O}(k^{p}_{\perp}/\omega^{p'})$ and ${\mathcal O}(k^{q}_{z}/\omega^{q'})$ with $p,q (p',q')<1 (>1)$. As a result, for $\omega \to \infty$, $F_{k,3}$ can be neglected compared to the leading order terms $F_{k,1}$.}\\

For type-I mWSM, one can keep in mind the fact that $b$ remains always positive. 
The total contribution from the Fermi surface is given by
\begin{eqnarray} \label{condfs3}
&& \sigma_{xy}^{FS (I) }\approx -
\frac{e^2 {n \alpha^{2-2/n}_n} }{4\pi^2 } \Bigg[
(\mu- C \Delta_n)\Big[\frac{v}{C^2}\ln \Big(\frac{v+C}{v-C}\Big) -2\Big] \nonumber \\
&+& 
 {\beta'_2 a(M) }\Big( \mu^{\frac{2M}{n}-3}
\frac{2 v (\mu- C \Delta_n) }{v^2-C^2} \nonumber \\ 
&-& 
{\Big(\frac{2M}{n}-3\Big) v^2 \mu^{\frac{2M}{n}-5} } 
\Big(\frac{\mu - C \Delta_n}{v^2-C^2} \Big)^3 (3C^2 +v^2) \Big) \nonumber \\
&+ & \beta'_3\{ M \to 2 M \} \Bigg]
\end{eqnarray}
with 
\begin{equation}
 a(M)=\frac{\Gamma(\frac{M}{n}+2 ) }{ (\frac{2M}{n}+2)
(\frac{2M}{n}-3) \Gamma(\frac{M}{n}+1)  }
\end{equation}
with $M=1$. Therefore, the leading contribution is not just given by 
${n \alpha^{2-2/n}_n}$ multiplied to $n=1$ contribution. In this first term $\mu$
gets renormalized by  $\mu-C \Delta_n$ while $\Delta_n$ depends on topological charge $n$.
The other 
sub-leading order terms are of order $1/\omega$. The multi Weyl nature 
thus imprints its effect on the Fermi surface part of the optical 
conductivity. We can write a closed form expression 
for $v\gg |C|$ as follows,
\begin{eqnarray} \label{condfs4}
&&\sigma_{xy}^{FS(I)}  =  n \frac{\alpha^{2-2/n}_n}{v}\cdot \frac{e^2}{4 \pi^2} 
\Bigg[  \frac{C(\mu-C\Delta_n)}{6 v^2}  \nonumber \\
&+& 4 \beta_2^\prime a(M) 
\Big( \frac{(2\mu)^{2/n-2}}{v}+ \frac{(2/n-3)\mu^{2/n-2}}{v} \Big) \nonumber \\
& + & 
{4 \beta_3^\prime a(2M)} \Big( \frac{2\mu^{4/n-2}}{v} + 
\frac{(4/n-3) \mu^{4/n-2}}{v}\Big)\Bigg]
\end{eqnarray}

Therefore, total conductivity of type-I mWSM for a given $n$ is expressed as 
\begin{eqnarray}
\label{condtot}
&&\sigma_{xy}^{I}(n)= n \frac{e^2}{4 \pi^2} \frac{ \alpha^{2-2/n}_n}{v}
\Bigg[ (Q+\Delta_n)+ C \Big( \frac{\mu-C\Delta_n}{6 v^2} \Big) \nonumber \\
&+& 4\beta''_2 a(M) \mu^{2/n-2}+
4\beta''_3 a(2M) \mu^{4/n-2}\Bigg]
\end{eqnarray}
with $\beta''_2 =\beta'_2 (\frac{2}{v}+\frac{2/n-3}{v})$ and $\beta''_3 =\beta'_3 (\frac{2}{v}+\frac{4/n-3}{v})$. 
This helps us to write the anomalous  thermal Hall conductivity $K_{xy}^I$ and Nernst conductivity 
$\alpha_{xy}^I$ respectively for the type-I mWSMs as ,
\begin{eqnarray} \label{therm1}
K_{xy}^I(n)  &=&   \frac{\pi^2}{3 e^2} k_B^2 T \sigma_{xy}^{I}  \nonumber  \\
& =&  n \frac{T k_B^2}{12} \frac{\alpha^{2-2/n}_n}{v}
\Bigg[ (Q+\Delta_n) - C \Big( \frac{\mu-C\Delta_n}{6 v^2} \Big)  \nonumber  \\
& + & 
4 \beta_2^{\prime \prime} a(M) 
 \cdot  \mu^{2/n-2} 
 +  4 \beta_3^{\prime \prime} a(2M)  \cdot  \mu^{4/n-2} \Bigg]\nonumber \\
\end{eqnarray}
One can find
\begin{eqnarray} \label{nerst1}
\alpha_{xy}^I (n)  &=&    \frac{\pi^2}{3 e^2} k_B^2  T \frac{d \sigma_{xy}^{I}}{d \mu}  \nonumber \\
& = & n \frac{e k_B^2}{12} \cdot \frac{\alpha^{2-2/n}_n}{v}\Bigg[- \frac{C}{6 v^2}
+ 4 \beta_2^{\prime \prime} a(M)   \Big(\frac{2}{n}-2\Big)   \nonumber \\
&& \mu ^{2/n-3} +4 \beta_3^{\prime \prime}  a(2M)   \Big(\frac{4}{n}-2\Big) \mu ^{4/n-3} \Bigg]
\end{eqnarray}
One can now easily derive the expressions for $\sigma_{xy}^{I}$, $K_{xy}^I $
and $\alpha_{xy}^I$ for $n=2$ by considering 
$\beta'_3=0$. Comments on the new results
for $n=2$ and $n=3$ and their characteristic  dissimilarities  from the $n=1$ case are now in order. In general, 
non-linear $\mu$ dependence comes from order $1/\omega$  term
in $n>1$ multi Weyl case while 
the linear $\mu$ dependence term only appear for $n=1$.  \\



\begin{figure*}[ht] 
\centering
\begin{subfigure}[t]{.45\textwidth}
\includegraphics[scale = 0.6]{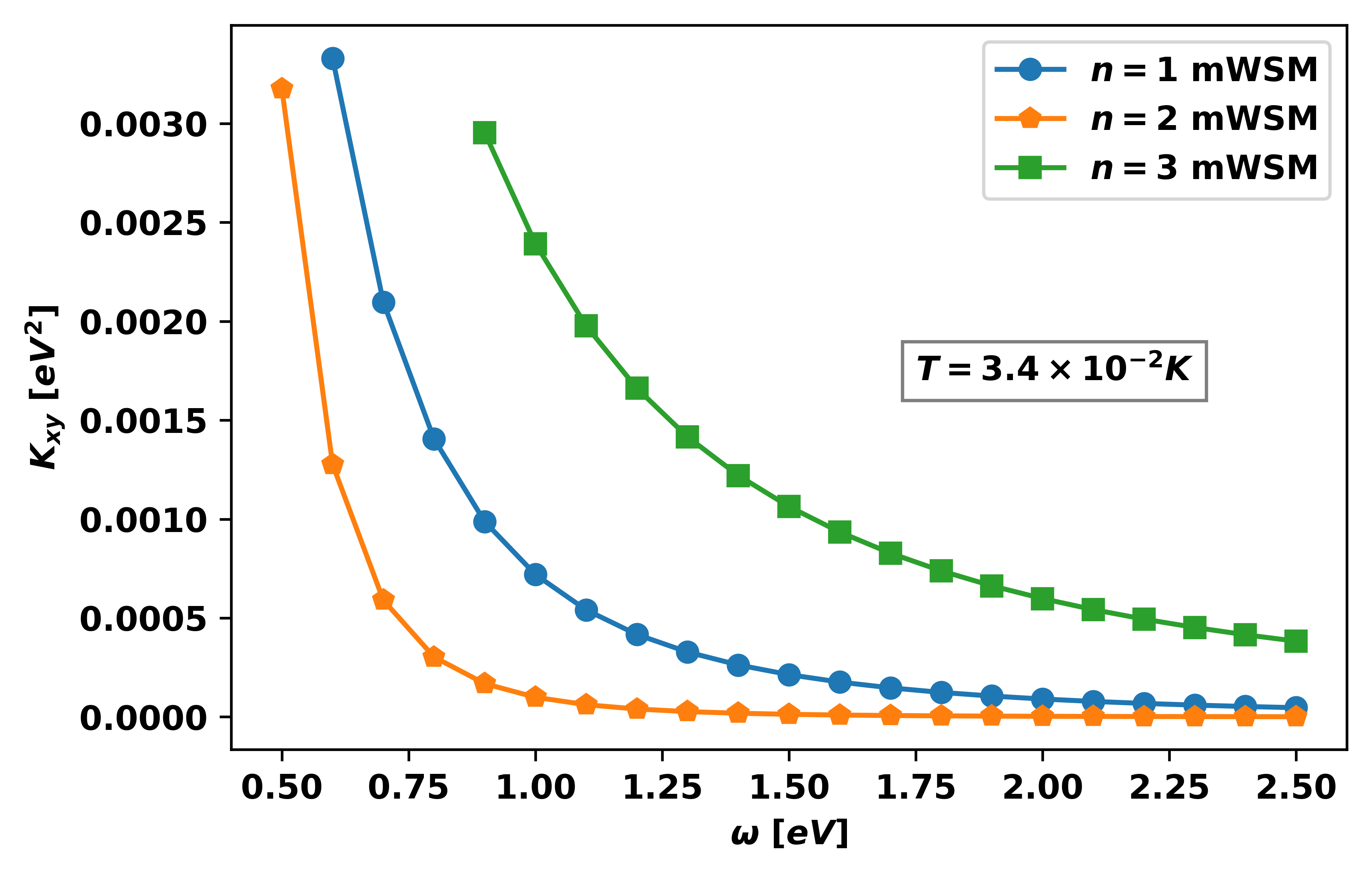}
\caption{}
\end{subfigure}
\hfill
\begin{subfigure}[t]{.45\textwidth}
\includegraphics[scale=0.6]{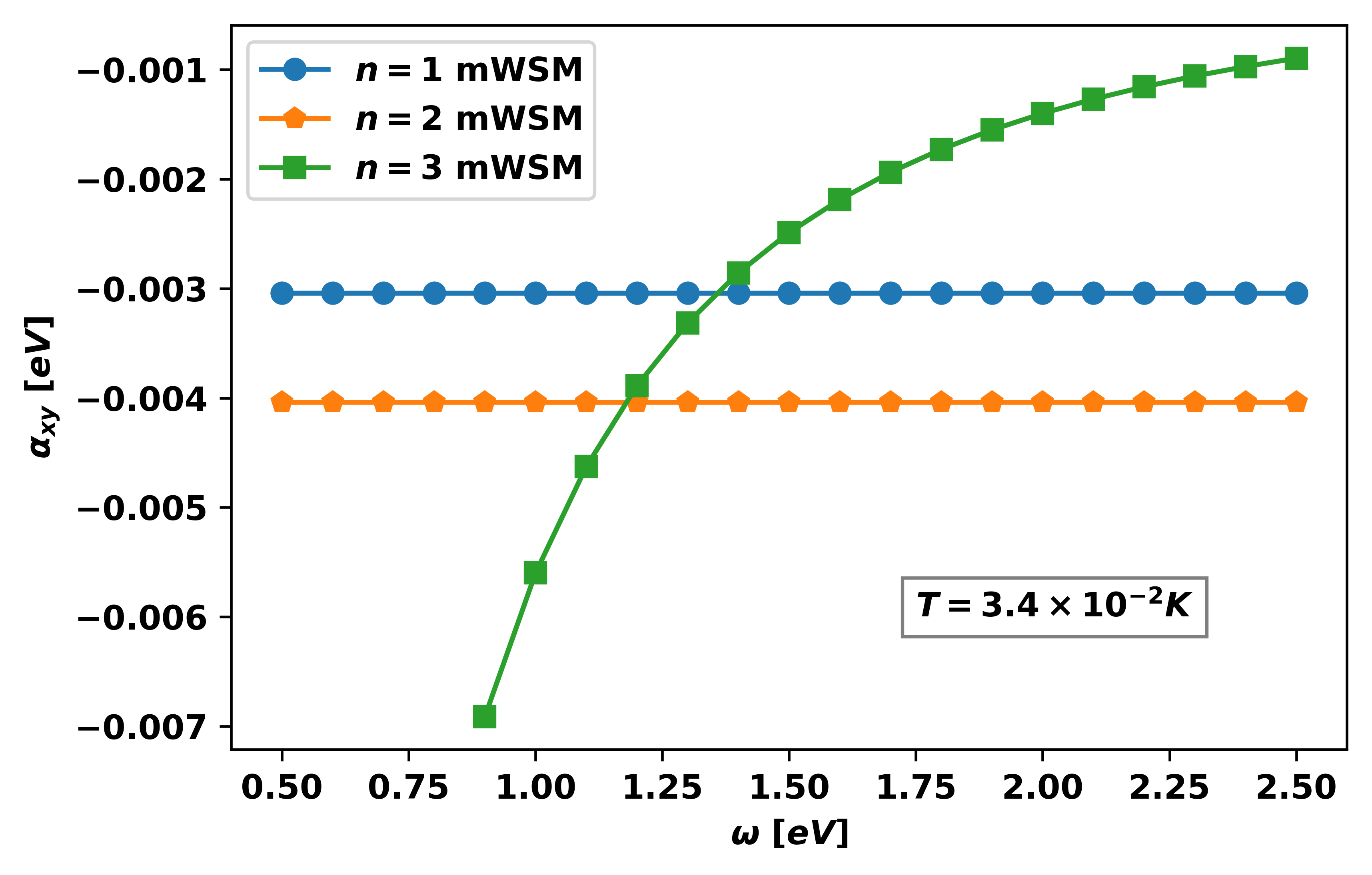}
\caption{}
\end{subfigure}
\caption{Type-I WSM: (a) Variation of thermal anomalous Hall conductivity with optical frequency, for three different values of the monopole charge. (b) Variation of anomalous Nernst conductivity with optical frequency, for three different values of the monopole charge.
\textcolor{black}{ The values of the various parameters are specified in Natural units as follows: $v_F = 0.005$, $\alpha_1 = v_F$, $\alpha_2 = 0.00012~ \textrm{eV}^{-1}$, $\alpha_3= 0.00012~ \textrm{eV}^{-2}$, $E_0 = \omega A_0 = 1000.0~ \textrm{eV}^2$, $ C =0.1$, $\mu = 1.0$ eV, $Q= 2.0$ eV,} \textcolor{black}{ and $T = 3.4 \times 10^{-2} $K. }}
\label{fig1_typeI}
\end{figure*}


\begin{figure} [ht] 
\centering
\includegraphics[scale = 0.6]{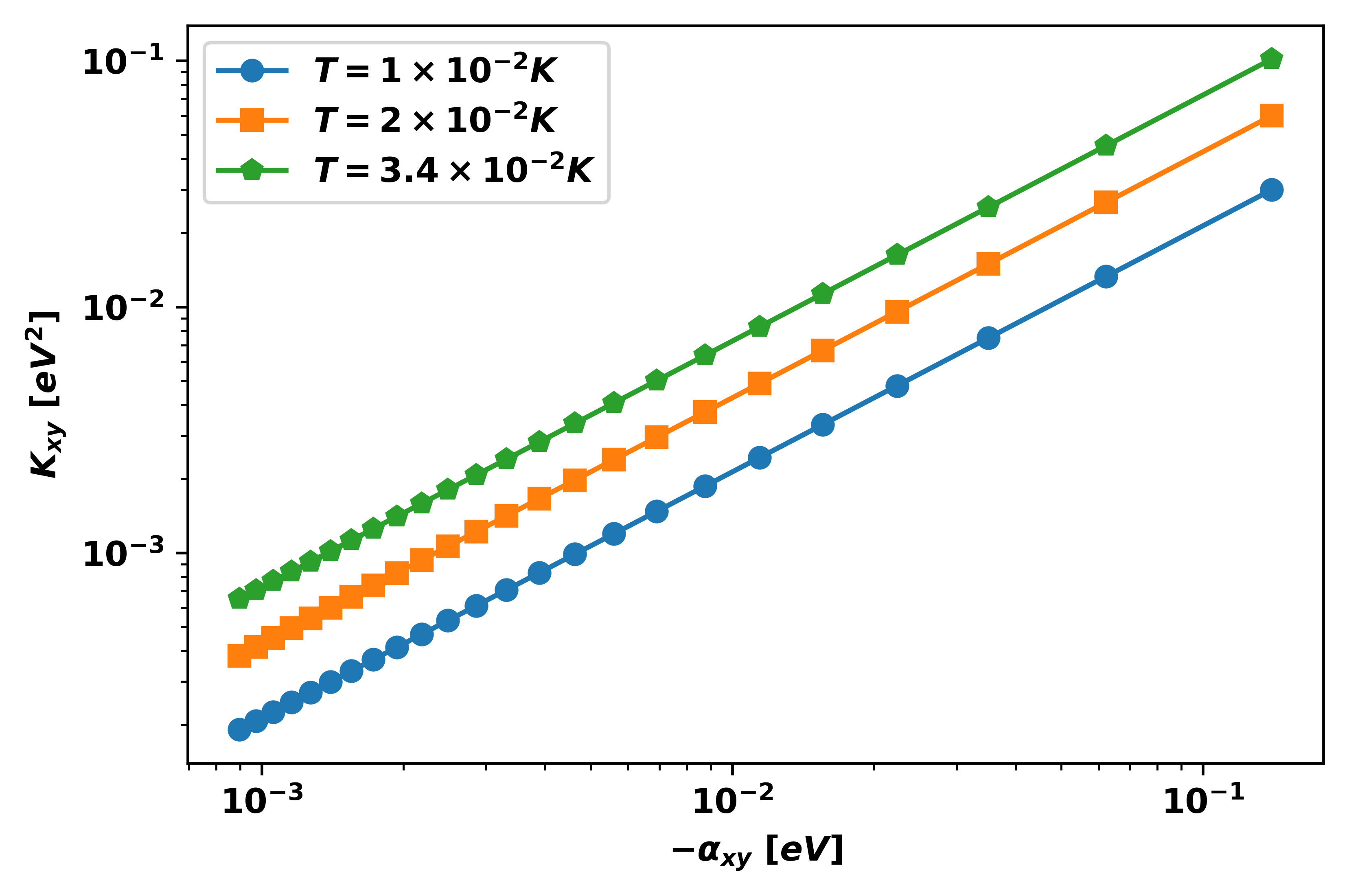}
\caption{Type-I WSM: Variation of thermal anomalous Hall conductivity with Anomalous Nernst conductivity, for $n=3$ mWSM. The frequency range sampled is $0.50$eV - $2.50$eV \textcolor{black}{The temperature values sampled are $T = 1 \times 10^{-2}, 2 \times 10^{-2}, 3.4 \times 10^{-2}$ K. }\textcolor{black}{ The values of the other various parameters are the same as Fig. \ref{fig1_typeI}. }}
\label{fig2_typeI}
\end{figure}

Let us now explore the thermal responses for the type-II case of mWSM where sign of 
$k_{\perp}$ momentum cut-off $b$ depends on $k_z$. Handling of 
the $k_{\perp}$ integral requires extra care  as ${\rm sgn}(b)$ becomes 
$+$ ($-$), depending on $k_z$ being $-(+)$. 
$|C|\gg v$, refers to the fact $v^2k_z^2- ( C k_z+sC \Delta_n-\mu)^2 <0$. Keeping 
in mind the $k_{\perp}$ integral, we find 
\begin{eqnarray} \label{condfstype2}
&& \sigma_{xy}^{II} (n) =  n \frac{e^2}{4 \pi^2} \frac{\alpha^{2-2/n}_n}{v} 
\Bigg[ ( \Delta_n +Q) \bigg(-1+ \frac{v}{C}\bigg) \nonumber \\
&-&\frac{v(C\Delta_n-\mu)}{C^2}
{\rm ln} \Big[\frac{C^2 \Lambda}{v(C\Delta_n-\mu)}\Big]  + 
 \beta_2^\prime a(M)\nonumber \\
&&
\bigg(\mu^{\frac{2M}{n}-3 }
\Big(\frac{2\Delta_n v}{C}- 2Q\Big) +\Big(\frac{2M}{n}-3\Big) \frac{C \mu^{\frac{2M}{n}-4}}{2} \nonumber \\
&&
\Big(\frac{4\Delta^2_n v}{C}- 2 \Lambda^2 -2 Q^2\Big) \bigg) 
+  \beta_3^\prime \{ M \to 2M\} \Bigg]\nonumber \\
\end{eqnarray}
with $M=1$.  The remarkable point to note here is that the momentum cut-off $\Lambda$  shows up algebraically in the Fermi surface 
contribution. However, this is accompanied with the sub-leading term  ${\mathcal O}(1/\omega)$. This is indeed a new feature for the 
anisotropic character of the dispersion in type-II mWSMs. In type-II single WSMs, the momentum cut-off can only appear logarithmically.\\

Using the results obtained above, we write the anomalous thermal Hall conductivity for type-II mWSMs:
\begin{eqnarray} \label{therm2}
&& K_{xy}^{II}(n) =   n T \frac{k_B^2}{12} {\alpha_n^{2-2/n}}
\Bigg[ (\Delta_n+Q) \bigg( \frac{v}{C}-1\bigg) \nonumber \\
&-& \frac{v}{C^2}(C\Delta_n-\mu) \ln \Big[ \frac{C^2 \Lambda}{v(C\Delta_n-\mu)} \Big] 
 +  \beta_2^\prime   a(M) \nonumber \\
&&\bigg( \mu^{2M/n-3} a_2(M) +  \mu^{2M/n-4} a_3(M) \bigg) \nonumber \\
 &+& \beta_3^\prime \{M \to 2M\} \Bigg
] 
\end{eqnarray}
with $a_2(M)=(2 \Delta_n v/C - 2 Q)$, $a_3(M)=C(2M/n -3)
(4 \Delta_n^2 v/C - 2 \Lambda^2- 2 Q^2)$ and $M=1$. 
On the other hand, the 
Nernst conductivity is given by 
\begin{eqnarray} \label{nernst2}
&&\alpha_{xy}^{II} (n) =   n  e \frac{k_B^2 \alpha_n^{2- 2/n}}{12 } \Bigg [
\frac{1}{C^2} \bigg[ - 1 + \ln \Big[ \frac{C^2 \Lambda}{ v(C\Delta_n-\mu)} \Big] \bigg]   \nonumber \\
& + &   \beta_2^\prime 
 a(M) 
\bigg( \Big(\frac{2M}{n}-3 \Big)\mu^{2M/n-4} a_2(M)  \nonumber \\
&+& \Big(\frac{2M}{n}-4 \Big) \mu^{2M/n-5} a_3(M) \bigg)
+  \beta_3^\prime  \{M \to 2M \} \Bigg] \nonumber \\
\end{eqnarray}
One can easily obtain the $n=2$ results by considering $\beta'_3=0$.



\begin{figure*} [ht] 
\centering
\begin{subfigure}[t]{.4\textwidth}
\includegraphics[scale = 0.54]{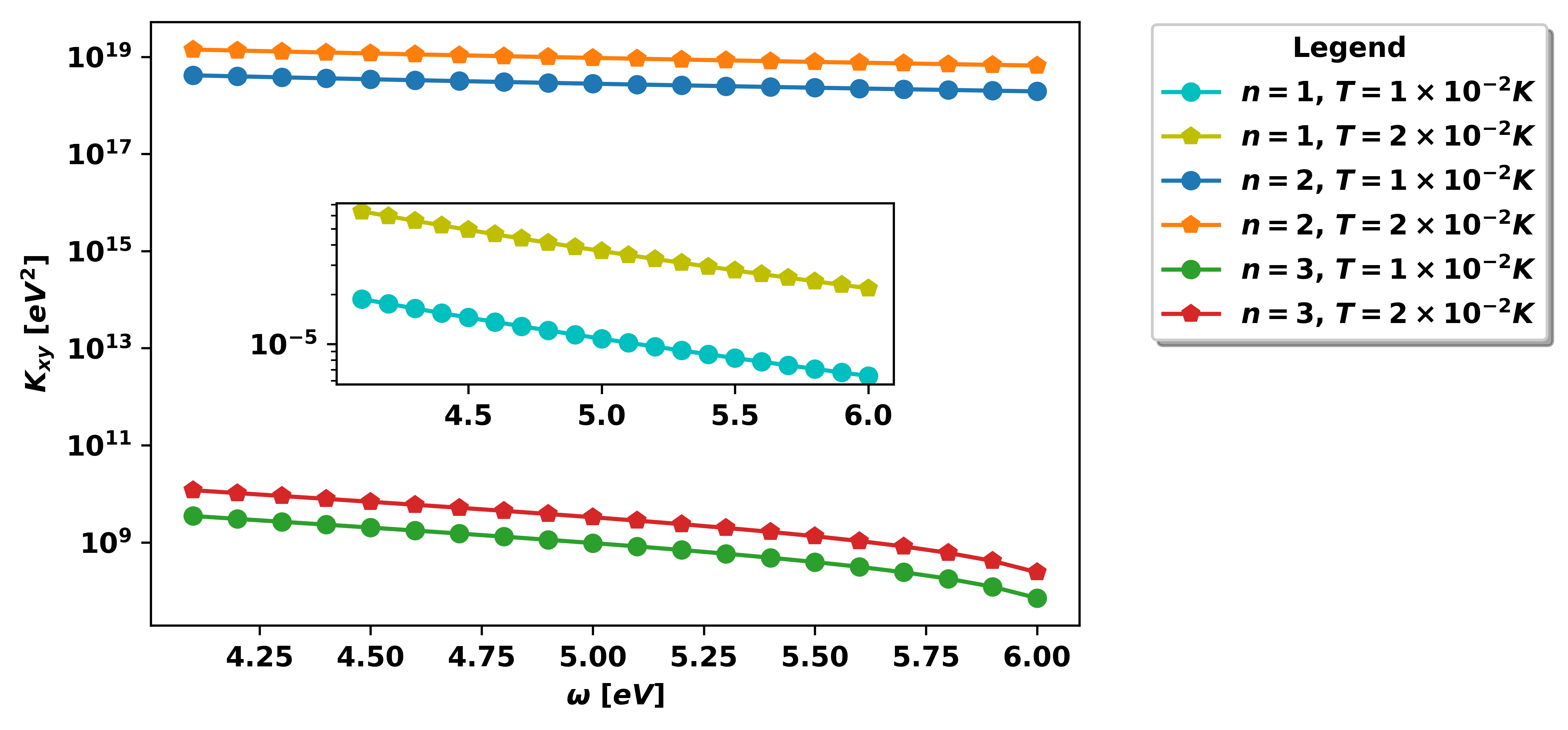}
\caption{}
\end{subfigure}
\hfill
\begin{subfigure}[t]{.4\textwidth}
\includegraphics[scale=0.53]{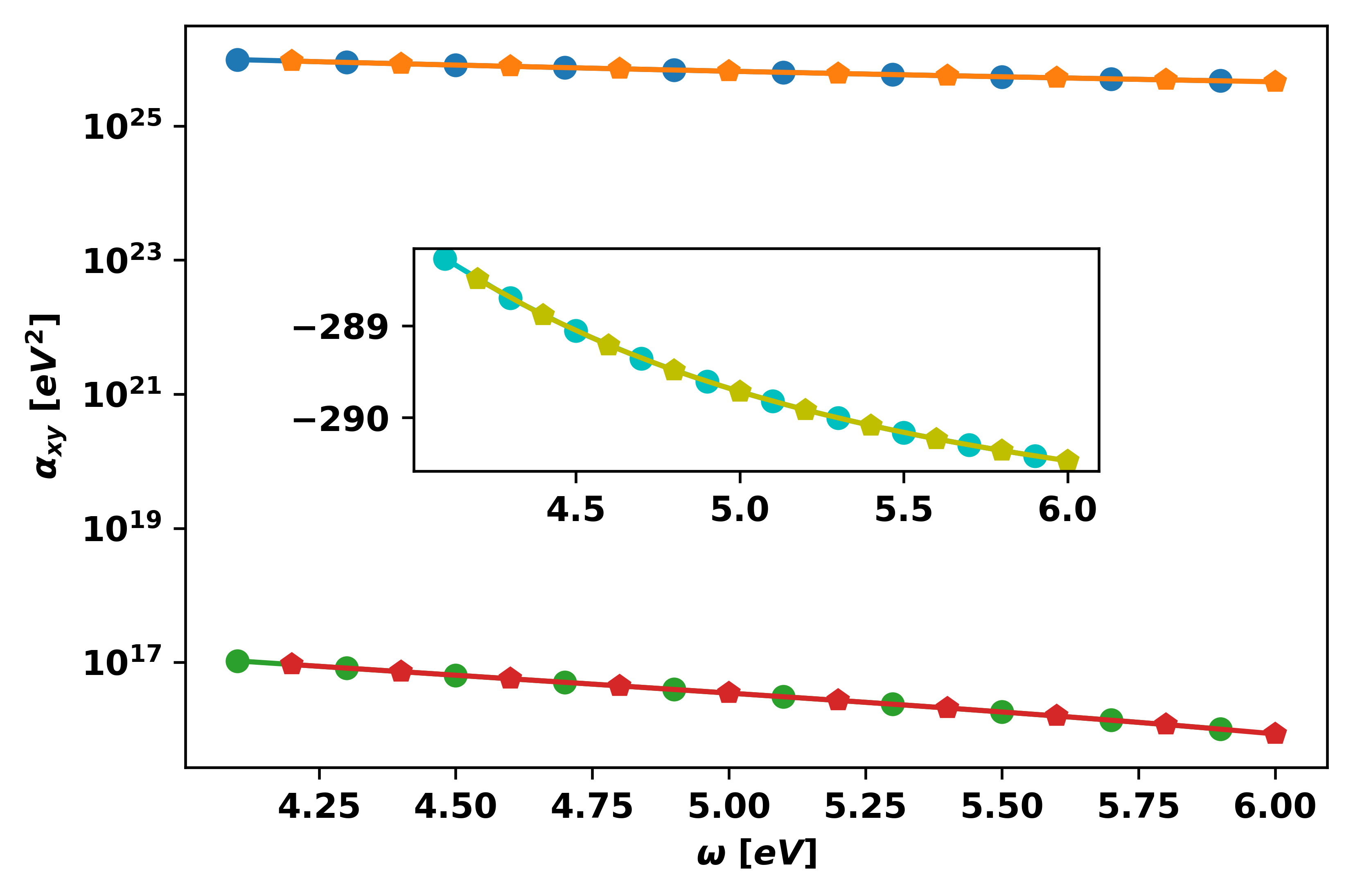}
\caption{}
\end{subfigure}
\caption{Type-II WSM: (a) Variation of thermal anomalous Hall conductivity with optical frequency, for three different values of the monopole charge. (b) Variation of anomalous Nernst conductivity with optical frequency, for three different values of the monopole charge.
\textcolor{black}{The plot (b) shows a strong overlap between the curves for fixed $n$ as a function of temperature. To highlight this issue, the data points sampled for overlapping curves are at distinct values of frequency.} \textcolor{black}{ The values of the various parameters are specified in Natural units as follows: $v_F = 0.005$, $\alpha_1 = v_F$, $\alpha_2 = 0.00012~ \textrm{eV}^{-1}$, $\alpha_3= 0.00012 ~ \textrm{eV}^{-2}$, $E_0 = \omega A_0= 1000.0 ~\textrm{eV}^2$, $ C =0.1$, $\mu = 1.0$ eV, $Q= 2.0$  eV, $\Lambda = 900.0$ eV, and $T = 1 \times 10^{-2}, 2 \times 10^{-2}$ K. }
}
\label{fig2_typeII}
\end{figure*}
	


\begin{figure*}[ht]
\centering
\begin{subfigure}[t]{.45\textwidth}
\includegraphics[scale = 0.6]{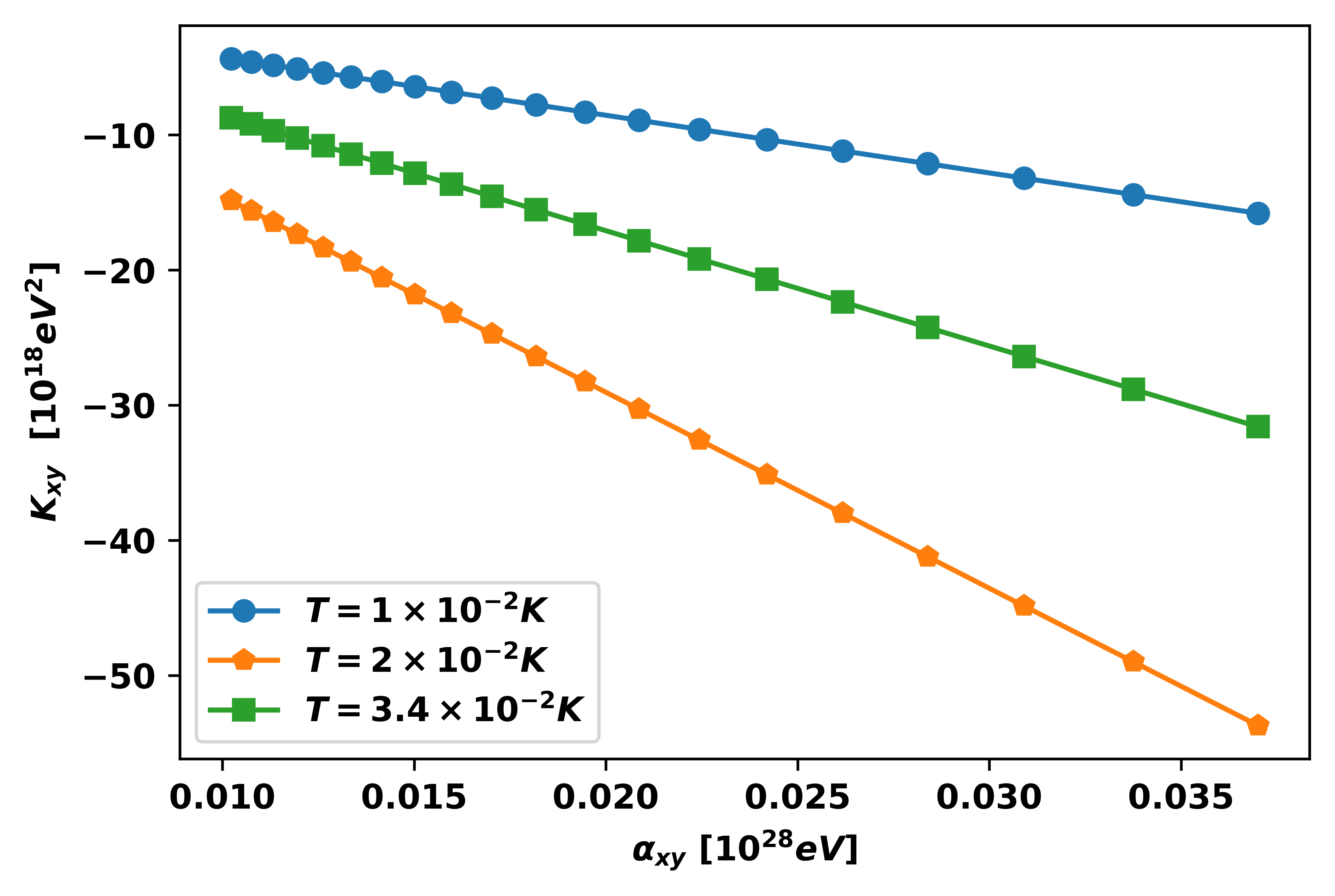}
\caption{}
\end{subfigure}
\hfill
\begin{subfigure}[t]{.45\textwidth}
\includegraphics[scale=0.6]{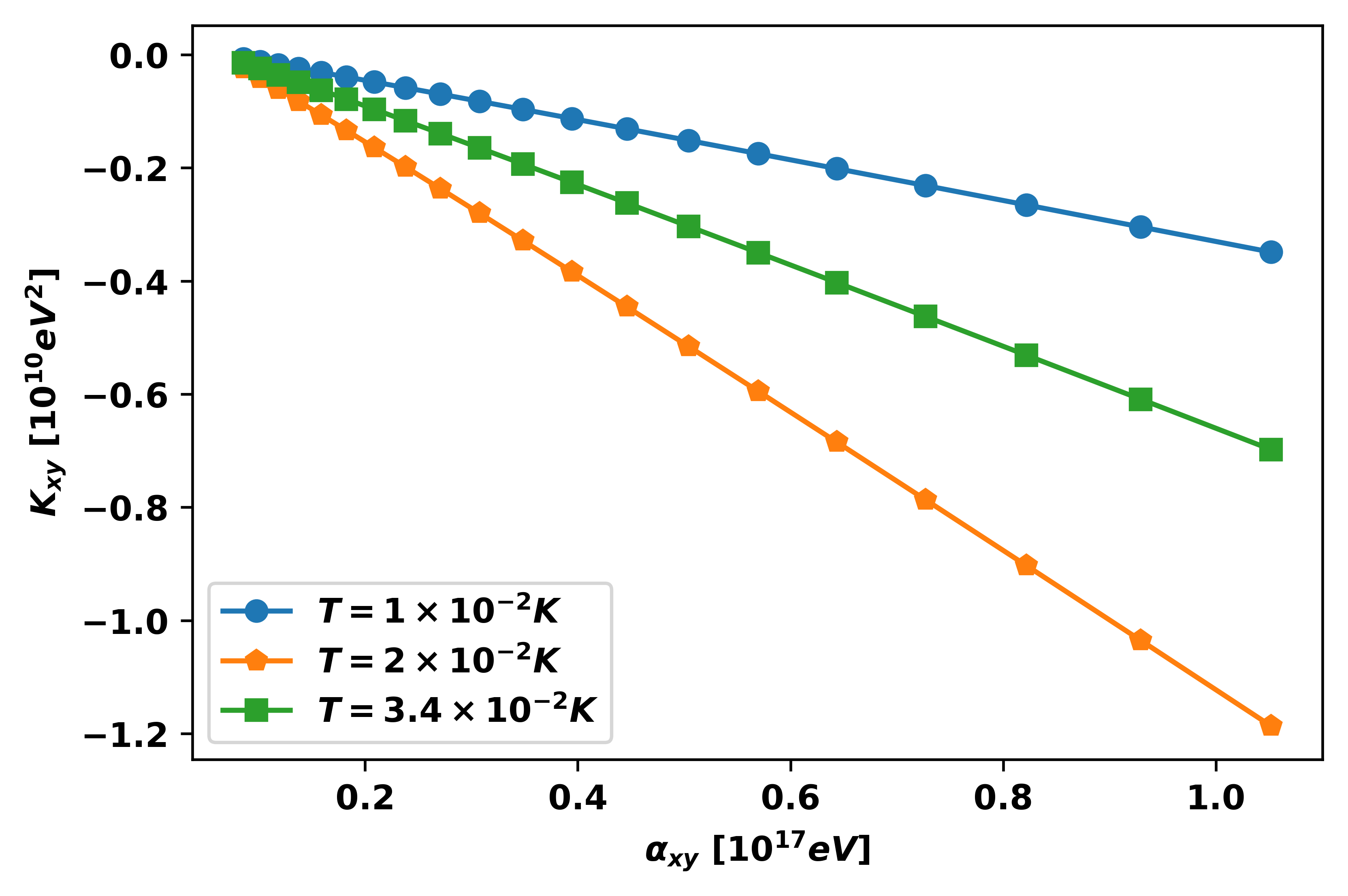}
\caption{}
\end{subfigure}
\caption{Type-II WSM: (a) Variation of thermal anomalous Hall conductivity with Anomalous Nernst conductivity, for $n=2$ mWSM. (b) Variation of thermal anomalous Hall conductivity with Anomalous Nernst conductivity, for $n=3$ mWSM. For both cases, the frequency range sampled is $4.1$eV - $6$eV.
\textcolor{black}{The temperature values sampled are $T = 1 \times 10^{-2}, 2 \times 10^{-2}, 3.4 \times 10^{-2}$ K.} \textcolor{black}{The values of the other various parameters are the same as Fig. \ref{fig2_typeII}.}
}
\label{fig3}
\end{figure*}
	
\section{Discussion of Results} \label{DR}

We now discuss some important aspects of our findings on the distinguishing transport features of type-I and type-II mWSMs.
\textcolor{black}{  First of all, we emphasize on 
the significant results that show the characteristically different features of  the effective chemical potential $\mu$ and the cut off $\Lambda$ for two different types of mWSMs. We also narrate  the key roles played by the topological charge $n$  and the  tilt $C$ in the thermo-electric transport properties of mWSMs in its two counterparts. 
We note that dispersion becomes anisotropic due to the multi Weyl nature;  tilt can additionally make it anisotropic in the tilt direction.  For mWSMs,  the Fermi surface becomes distorted circle or cone depending on the tilt in the static limit.  
The distribution of chiral Weyl fermions also takes part importantly in transport. Floquet driving  can  lead to further complicated deformation of the static Fermi surface. Moreover, it can influence the  distribution of chiral Weyl fermions in the electron and hole pockets. 
Therefore, Floquet transport can 
noticeably be altered upon the introduction of the tilt. In terms of the physical parameters, the differences in transport are clerly visible that are originated from the nature of the Fermi surface. Having qualitatively analyzed the differences, we below present their quantitative  
nature.}

 It is to be noted that $\Lambda^2$ is associated with $\mu^{2M/n-4}$ and $\mu^{2M/n-5}$, (with $M=1,2$) for optical  Hall conductivity, and Nernst conductivity in case of type-II mWSMs, respectively. Therefore, the  transport properties in this phase are heavily influenced by the coupling of $\mu$
and $ \Lambda$. This is contrary to the type-I mWSM where only $\mu$ can affect the transport in addition to the driving field; 
$\Lambda$ does not appear in the transport coefficients. For type-II single WSM, a purely logarithmic cut-off dependence is only observed. Hence, the anisotropy in the tilted dispersion non-trivially couples with the field parameters to generate the unusual cutoff dependence. The shape of the Fermi pockets for type-II mWSMs is very different from type-I mWSM as it evident from the cut-off dependence of transport coefficients. Notably, in case of irradiated tilted mWSMs, the topological charge imprints its effect not only in a simple multiplicative fashion but also in a much more fundamental way, by coupling to the tilt dependent effective chemical potential, where $\Lambda$ appears algebraically. This algebraic cut-off dependent term is associated with the additional corrections of ${\mathcal O}(1/\omega)$. The leading order term in the off-diagonal conductivity is given by $n$  times the single Weyl result; here, the anisotropic nature of the dispersion is partially encoded in the renormalized chemical potential $\mu \to \mu -C \Delta_n$, where  $\Delta_n={\mathcal O}(A_0^{2n}/\omega)$. The effective chemical potential is also dependent on the frequency of the driving potential and the monopole charge. \\

Having discussed the implication of cut-off, we here investigate the non-linear $\mu$ dependence that arises in the conductivity tensor, besides the effective $\mu$. In type-I mWSMs, considering $v\gg |C|$, the vacuum contribution $\sigma_{xy}^I$ associated with $\beta'_{2,3}$ term becomes decreasing function of $\mu$ for both for $n=2$ and $n=3$; $\beta'_2$ term decays inversely (as $\mu^{-1}$)  for $n=2$  and $\beta'_{2,(3)}$ decays non-linearly $\mu^{-4/3}(\mu^{-2/3})$ for $n=3$. The Nernst conductivity on the other hand, goes as $\mu^{-2}$ for $n=2$ and for $n=3$, it becomes decreasing function of $\mu$ (as $\mu^{-7/3}$ and $\mu^{-5/3}$). In type-II mWSMs, considering $ |C|\gg v$, the vacuum contribution $\sigma_{xy}^{II}$ associated with $\beta'_{2,3} \Lambda^2$ term becomes decreasing function of $\mu$ for both the $n=2$ and $n=3$ cases. We note that the sub-leading correction decays more rapidly with $\mu$ for type-II as compared to type-I mWSMs. 
{ In particular, the cut-off independent contributions asscociated with 
$\beta'_{2}$ term vary as $\mu^{-2}$ and $\mu^{-3}$ for $n=2$. While for $n=3$, these contributions associated with 
$\beta'_{2,(3)}$ term  go as $\mu^{-7/3}$ and $\mu^{-10/3}$ ($\mu^{-5/3}$ and $\mu^{-8/3}$).} The Nernst conductivity in this regime becomes strongly decreasing function of $\mu$ for both $n=2$ and $n=3$ with the lowest power as $\mu^{-3}$ and $\mu^{-8/3}$, respectively.
\\

After investigating the transport behavior analytically, we below illustrate them 
as a function of driving frequency
to analyze some salient qualitative features. We note that our aim is to pictorially differentiate the type-I from type-II mWSM based on our low-energy model. Hence, at the outset, we confess that certain lattice 
effects might not be captured following our analysis. However,  our study  uncovers some trends which we believe can be probed in real materials. \\

We now discuss  the transport coefficients for type-I mWSMs as shown in Fig.~\ref{fig1_typeI}(a) for thermal Hall conductivity and Fig.~\ref{fig1_typeI}(b) for Nernst conductivity. We here depict the high frequency behavior of $K_{xy}$ and $\alpha_{xy}$, 
calculated using Eq.~(\ref{therm1}) and Eq.~(\ref{nerst1}), respectively. Noticeably the response from the external field for a general $n>1$ mWSM is not related to $n=1$ single WSM by 
a simple multiplicative factor. This is also very clearly evident from the variation of $K_{xy}$ and $\alpha_{xy}$ with driving frequency $\omega$. The sub-leading terms play an important role due to the fact that the 
chemical potential $\mu$ gets non-trivially coupled to 
the frequency; these terms are associated with the factors $\beta'_2$, $\beta'_3$. The important point to note here is that $K_{xy}$
decreases and eventually saturates with optical frequency $\omega$; while $|\alpha_{xy}|$ remains unchanged with $\omega$ for $n=2$. In the case with $n=3$, $|\alpha_{xy}|$ increases followed by a saturation at sufficiently large frequency.
{\color{black} We note that even though $\beta''_2=\beta''_3=0$
for both $n=1$ and $n=2$, $K_{xy}$ depends on $\omega$ as first two terms in Eq.~(\ref{therm1}) encompass the factor $\Delta_n$. The $\omega$-independent nature of $\alpha_{xy}^{n=1}$ and $\alpha_{xy}^{n=2}$ stems from the fact that $\beta''_2=\beta''_3=0$ in the leading order; the  first term in  
Eq.~(\ref{nerst1}) does not depend on $\omega$. $\beta''_2,\beta''_3 \ne 0$ that result in $\omega$-dependent behavior of $\alpha_{xy}^{n=3}$.
The absence and lower degree of anisotropy can thus lead to $\omega$-independent nature of 
$\alpha_{xy}^{n=1}$ and $\alpha_{xy}^{n=2}$, respectively; 
substantial amount of anisotropy can significantly modify 
the light induced transport as observed in $\alpha_{xy}^{n=3}$.
However, the crossing of 
$\alpha_{xy}^{n=3}$ with $\alpha_{xy}^{n=1}$ and $\alpha_{xy}^{n=2}$ might be restricted to the leading order and higher order correction can be frequency dependent that we do not calculate here. We can comment that one needs to investigate the lattice model to get the complete picture.}

\textcolor{black}{It is now important to analyze the behavior of 
$K_{xy}$ as a function of $\alpha_{xy}$ that could be useful from the experimental perspective. One can understand that  $K_{xy}$ and $\alpha_{xy}$ 
behave in an independent manner for $n=1$ and $2$ as $K_{xy}$ 
depends on $\omega$ while $\alpha_{xy}$ does not.
 Interestingly, we see that this no longer holds for $n = 3$ and we plot this in Fig. \ref{fig2_typeI}. Here,  $K_{xy}$ increases with $|\alpha_{xy}|$. 
 A qualitative change in the transport character is observed with the increase in the degree of anisotropy, characterized by $n$.}

Similarly, for type-II mWSMs,  we depict the behavior of $K_{xy}$, obtained from Eq.~(\ref{therm2}), in Fig.~\ref{fig2_typeII} (a) and  $\alpha_{xy}$, obtained from Eq.~(\ref{nernst2}), in Fig.~\ref{fig2_typeII} (b), respectively. One can find here for type-II mWSM, unlike the type-I mWSM, that $K_{xy}$  and $\alpha_{xy}$ both  decrease with $\omega$. This may be due to the fact that they are influenced by the quadratic momentum cutoff $\Lambda^2$ dependent sub-leading term in addition to the terms containing the function $f(\mu, \omega, n)$.
\textcolor{black}{
We note in the sufficiently large frequency regime
that  conductivities for type-I tripple WSM are at a higher magnitude  as compared to single and double WSM, while this is not the case for type-II. The responses from type-II double  WSM acquire maximum  value.  For type-I, conductivities of double WSM become lowest in the sufficiently large frequency regime; 
in contrast, the conductivities for type-II single WSM becomes vanishingly small as shown in the insets of Fig.~\ref{fig2_typeII} (a) and (b)}.
\textcolor{black}{We note that $\alpha_{xy}$ behaves identically with temperature 
for type-II single WSMs and mWSMs as shown in Fig.~\ref{fig2_typeII} (b).}
Having investigated anomalous thermal Hall and Nernst coefficients for a range of physically viable parameter such as $\omega$, we show that type-I and type-II mWSMs can be  qualitatively distinguished in terms of their transport behavior.   \\

Now we shall focus on the role of the topological charge $n$ in different transport properties. For that, we plot $K_{xy}$ as a function of $\alpha_{xy}$ for $n=2$ in Fig.~\ref{fig3}(a), and for $n=3$ in Fig.~\ref{fig3}(b). It is known that $K_{xy}$ and $\alpha_{xy}$ share a linear relationship \cite{menon18} for $n=1$, and we notice that this holds for $n=2$: This can be attributed to the fact the first sub-leading order term remains small for a given chemical potential. This no longer holds for $n=3$ as is evident from Fig.~\ref{fig3}(b) where the sub-leading order terms play a crucial role. Therefore, one finds a qualitative change in the transport character with $n$, as the degree of anisotropy enhances. \textcolor{black}{ A comparison between Fig. \ref{fig2_typeI} and Fig.~\ref{fig3}(b) suggests that $K_{xy}$
decreases with increasing $|\alpha_{xy}|$ for type-II while 
$K_{xy}$ increases with increasing $|\alpha_{xy}|$ for type-I.
Therefore, tilt can significantly modify the transport even for the irradiated mWSMs.} \\


\textcolor{black}{Having thoroughly investigated the transport coefficients type-I and type-II mWSMs, we would now like to comment
on the differences between these two phases in single WSM 
as far as the other magneto-transport conductivities are concerned. 
As a start,  planar Hall coefficients vary quadratically (linearly) for type-I (type-II) single WSMs \cite{nandy17_PRL}. The type-I  single WSMs can be differentiated from type-II while the anomalous Nernst and anomalous Hall conductivities  are studied \cite{saha18,sharma17}. The tilt also causes distinguishably different optical activities in Kerr and Faraday rotation as  compared to the non-tilted case  \cite{sonowal19,kargarian15}. Our study 
considering the low energy  irradiated  mWSM model
further strengthens the list of distinction between these two types of mWSMs. The distinct behavior coming from type-I and type-II single Weyl lattice models which do not suffer from any cut-off dependence can thus be related to the different cut-off characteristics as derived in low energy model. Therefore, the tilt even in the presence of anisotropy is able to influence the transport properties in a different manner as compared to non-tilted case. }\\


We shall now propose a relevant experimental setup where our predictions can be tested. One can have candidate double (HgCr$_2$Se$_4$) and triple WSM (Rb(MoTe)$_3$) materials as the samples. The Floquet driving can be realized by the conventional pump (strong beam)-probe (weak beam) optical set up where ultrafast electron dynamics of the samples are observed as a function of time delay between the arrival of pump and probe pulses. Recently, using polarized photons at mid-infrared  wavelengths, Floquet-Bloch states and photo-induced band gaps have been shown to be clearly visible in  time-and-angle-resolved photoemission spectroscopy \cite{wang13}. We believe that using similar arrangements with suitably chosen frequency ranges of pump laser, one can experimentally measure the transport properties derived here. One can also consider a non-optical substrate-terminal based closed circuit  measurement of Nernst conductivity and thermal Hall conductivity \cite{watzman18} The electric and heat current can be measured considering a mutually perpendicular arrangement of DC power source and thermocouple, respectively.\\

\section{conclusion}
\label{conclusion}

In this manuscript we have investigated the circularly polarized light (of amplitude $A_0$ and frequency $\omega$) induced contributions to the thermo-electric transport coefficients in  type-I and type-II mWSM with topological charge $n>1$ considering the low energy minimal model. Using the high frequency expansion ($\omega \to \infty$) and appropriately employing the non-equilibrium Floquet-Kubo formalism, where the energies and states of the Hamiltonian are replaced with the  quasi-energy and quasi-states of the effective Hamiltonian, we study the anomalous thermal Hall conductivity and Nernst conductivity. The effective Floquet Hamiltonian  suggests that  the  Weyl  nodes, separated by $Q$ in the momentum space for the static case, are further displaced by a distance $2\Delta_n\sim A_0^{2n}/\omega$. Importantly, the low energy Hamiltonian of Floquet mWSMs receive momentum dependent corrections in addition to the constant $A_0^2$ shift in the single $n=1$ Floquet WSMs. This results in a change in the effective Fermi surface which in turn leads to an array of non-trivial consequences for the transport coefficients. The leading order contribution varies linearly with the topological charge  and 
 the chemical potential $\mu$ is  renormalized to $\mu-C\Delta_n$. Therefore, the light induced transport phenomena in type-I, and type-II mWSMs become significantly different. In particular, one can show that optical conductivity increases with $A_0$ for type-I mWSMs, while it decreases with $A_0$ in the case of type-II mWSM. However, the leading order vacuum contribution to $\sigma_{xy}$
  remains topological, which we verify by calculating  
  the Berry curvature induced anomalous Hall conductivity.\\

Going beyond the leading order contrbution, we compute the effect of the momentum dependent correction term in the Fermi surface effects to the conductivity tensor. We find Floquet driving induced sub-leading contribution can show non-trivial algebraic dependence on the chemical potential $\mu$ as $\mu^{f(n)}$. Most surprisingly, unlike the case of type-II single WSMs, for type-II mWSMs, the Nernst and thermal Hall conductivity depends algebraically on the momentum cut-off. However, for type-I mWSMs, the Fermi surface contribution remain cut-off independent. On the other hand, it decays slowly for type-I mWSM as compared to type-II mWSM. Consequently, unlike the type-I single WSM, the Nernst conductivity for type-I mWSM depends on $\mu$. Combining all these, we graphically represent the variation of
 the total thermal Hall and Nernst conductivities as a function of the optical driving frequnecy by  evaluating the analytical expression numerically. These suggest that type-I and type-II mWSM exhibit distinct behavior while the multi Weyl nature can also be captured vividly. 
 \textcolor{black}{
This would directly connect our study with the possible future experiments.
In conjunction to the previous point, we discuss about the possible experimental measurements and setups of our analytical findings. 
Therefore, we believe that our work could motivate a plethora of studies in the related experimental and theoretical areas
dealing with driven WSMs.} \\

{\bf Acknowledgement.} {\it TN specially thanks  MPIPKS, Dresden, Germany, for providing the local hospitality.}



\end{document}